\documentclass[12pt]{spieman}  
\pdfoutput=1
\usepackage{amsmath,amsfonts,amssymb}
\usepackage{graphicx}
\usepackage{setspace}
\usepackage{tocloft}
\usepackage{subcaption}
\usepackage{float}
\usepackage{xcolor}
\usepackage{lineno}
\usepackage{soul}

\newcommand\elec{\mathrm{e^-}}

\title{Characterization of Low Light Performance of a CMOS Sensor for Ultraviolet Astronomical Applications}

\author[a]{Timothee Greffe}
\author[a]{Roger Smith}
\author[b]{Myles Sherman}
\author[b]{Fiona Harrison}
\author[b]{Hannah Earnshaw}
\author[b]{Brian Grefenstette}
\author[c]{John Hennessy}
\author[c]{Shouleh Nikzad}
\affil[a]{Caltech Optical Observatories, California Institute of Technology,  Pasadena, California, USA}
\affil[b]{Cahill Center for Astronomy and Astrophysics, California Institute of Technology,  Pasadena, California, USA}
\affil[c]{Jet Propulsion Laboratory, California Institute of Technology, Pasadena, California, USA}

\cftpagenumbersoff{figure}
\cftpagenumbersoff{table} 
\begin{document} 
\maketitle

\begin{abstract}
CMOS detectors offer many advantages over CCDs for optical and UV astronomical applications, especially in space where high radiation tolerance is required.     However, astronomical instruments are most often designed for low light-level observations demanding low dark current and read noise, good linearity and high dynamic range, characteristics that have not been widely demonstrated for CMOS imagers.  We report the performance, over temperatures from 140 - 240\,K, of a radiation hardened SRI 4K$\times$2K back-side illuminated CMOS image sensor with surface treatments that make it highly sensitive in blue and UV bands. After suppressing emission from glow sites resulting from defects in the engineering grade device examined in this work, a 0.077\,$\mathrm{m\elec/s}$ dark current floor is reached at 160\,K, rising to 1\,$\mathrm{m}\elec/\mathrm{s}$ at 184\,K, rivaling that of the best CCDs.  We examine the trade-off between readout speed and read noise, finding that 1.43\,$\elec$ median read noise is achieved using line-wise digital correlated double sampling at 700\,$\mathrm{kpix/s/ch}$ corresponding to a 1.5\,$\mathrm{s}$ readout time. The 15\,$\mathrm{k}\elec$ well capacity in high gain mode extends to 120\,$\mathrm{k}\elec$ in dual gain mode.  Continued collection of photo-generated charge during readout enables a further dynamic range extension beyond $10^6\,\elec$ effective well capacity with only 1\% loss of exposure efficiency by combining short and long exposures. A quadratic fit to correct for non-linearity reduces gain correction residuals from $1.5\%$ to $0.2\%$ in low gain mode and to $0.4\%$ in high gain mode. Cross-talk to adjacent pixels is only $0.4\%$ vertically, $0.6\%$ horizontally and $0.1\%$ diagonally. These characteristics plus the relatively large ($10\mathrm{\mu}\mathrm{m}$) pixel size, quasi 4-side buttability, electronic shutter and sub-array readout make this sensor an excellent choice for wide field astronomical imaging in space, even at FUV wavelengths where sky background is very low.

\end{abstract}

\keywords{CMOS,SRI mK$\times$nK, ultraviolet, dark current, glow suppression, dual-gain}

{\noindent \footnotesize\textbf{*}Timothee Greffe,  \linkable{tgreffe@caltech.edu} }

\begin{spacing}{2}   

\section{Introduction}
\label{sect:intro}  

For decades the detector of choice for ultraviolet (UV) space applications has been the microchannel plate (MCP) coupled to various kinds of 2D charge sensors. Space missions such as GALEX, Neil Gehrels Swift, and Hubble Space Telescope Cosmic Origins Spectrograph adopted MCP imagers because their sensitivity extends into the far-UV (FUV), while the wide bandgap of their photocathode depresses sensitivity to optical photons and allows them to be operated at room temperature. However, MCPs have relatively low quantum efficiency (QE) of only 10\% at $1500\mathrm{Å}$ [\citenum{10.1117/12.460034}] [\citenum{10.1117/12.2561753}] [\citenum{Nikzad:12}], require high voltages in space where electrical breakdown is a concern, and are damaged by over-illumination. The latter issue has restricted many astronomical UV telescopes from viewing fields containing bright stars, particularly in the FUV where over-illumination causes severe depletion of the photocathode.  

The drawbacks of MCPs have motivated the development of Silicon-based UV imagers (CCD and CMOS) for astronomy, where higher QE provides more sensitivity 
for a given telescope size, enabling powerful observatories on moderate-sized platforms.   The ability to view fields containing bright stars and regions where high dynamic range is required, such as the Galactic plane and Magellanic Clouds, is another significant scientific advantage, especially for wide-field of view (FoV) systems.    On the technical side, avoiding high voltages and being able to match pixel sizes to the specific application are additional positive factors.   For wide FoV applications, CCDs and CMOS detectors can be assembled in close butted mosaics with $\sim$85\% fill factor, supporting imaging applications on a larger scale than MCPs[\citenum{10.1117/12.2561753}].

The Ultraviolet Explorer (UVEX) mission [\citenum{arXiv:2111.15608}], which motivates the investigation documented here, requires two 12K$\times$12K focal planes, imaging in FUV($1390-1900\,\mathrm{Å}$) and NUV ($2030-2700\,\mathrm{Å}$) passbands at a plate scale of $1.02^{\prime\prime}$ per 10$\,\mathrm{\mu}\mathrm{m}$ pixel, and a high efficiency Long Slit Spectrometer covering the $1150-2650\,\mathrm{Å}$ wavelength range, at $0.4\,\mathrm{Å/pixel}$ sampling with a  single detector.  The sky in the FUV is so dark that even a small amount of readout noise dominates photon shot noise from sky in a typical 900s exposure, so the ideal sensor will have very low read noise ($\sim2\,\elec$) and low dark current ($<10^{-3}\,\elec/\mathrm{s}$) at temperatures readily accessible with radiative cooling (170K).

While  CCDs can meet some of these requirements at beginning-of-life, CMOS image sensors have very significant advantages [\citenum{10.1117/12.148585}].   Since charge is converted to voltage within the pixels, there are negligible charge transfer losses, a problem which is particularly severe in CCDs when background is low, and gets worse with radiation damage,  requiring complex post processing based on frequent recalibration as demonstrated on HST[\citenum{arXiv:1401.1151}][\citenum{arXiv:1009.4335}].  Also, CMOS imagers do not require a shutter.  While CCDs can also be operated without a shutter, the required subtraction of averaged parallel overscan lines adds shot noise and delivers poor cancellation of image smear in the presence of pointing jitter. 

In the CMOS architecture studied here, the  5-transistor pixel design (Figure \ref{Fig:PIXEL})  eliminates dead time between exposures.  Charge transfer to the sense node defines the end of one exposure and the start of the next, allowing charge collection to continue during readout without charge smearing.  We use Rolling Shutter mode which requires one frame time to clear charge at the start of a sequence of exposures. Subsequent readouts between exposures of the same scene do not represent any overhead.  An important application of this is to extend the dynamic range by combining consecutive short and long exposures, e.g. pairing 3 s and 300 s exposures extends dynamic range by 100 without dead time between exposures due to readout.

Another advantage of CMOS imagers is the flexibility they afford for subarray readout.
Since pixels are addressed directly, subarrays can be read at high cadence without disturbing signal integration on the remainder of the sensor.  This feature further extends the dynamic range to bright targets which would saturate in the normal frame time, and enables high time resolution studies of rapidly varying sources.   For space applications where guide stars are required, subarray readout can provide rapid centroiding of bright stars without interfering with the use of the rest of the device for deep science exposures.

\begin{figure}[H]
    \begin{center}
       \includegraphics[width=120mm]{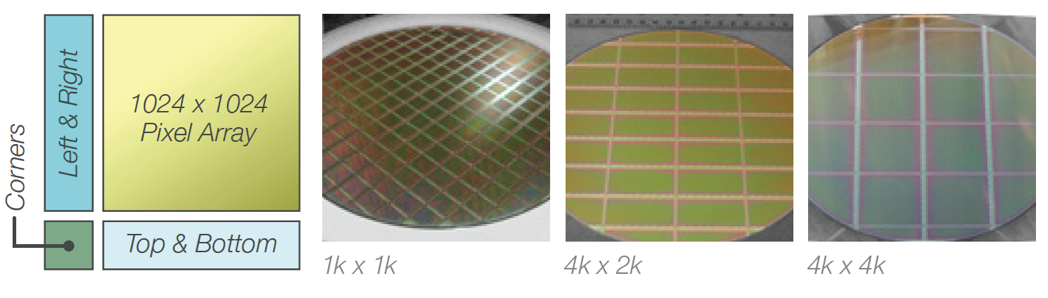}
    \end{center}
\caption{ The Mk$\times$Nk Imager reticle consists of four modular building blocks which are
photocomposed (“stitched”) to create custom
dimensions in 1k increments. The sizes shown have been successfully produced, for ground and space missions.} 
\label{Fig:SRIWafer}
\end{figure}

In this paper we report on the characterization of an SRI Nk$\times$Mk CMOS sensor[\citenum{10.1117/12.2189941}]  made by Sarnoff International for scientific imaging from space.  This radiation tolerant design (Table \ref{tab:RadiationTolerance}) has significant flight heritage (Parker Solar Probe, Solar Orbiter, SOLO HI EIS) [\citenum{2017AGUFMSH23D2693P}] [\citenum{10.1117/12.2027657}].  It can be fabricated in 1k increments up to 4k$\times$4k from the same reticles (Figure \ref{Fig:SRIWafer}) and the compact address logic allows 4k$\times$4k tiles to be mosaicked with high fill factor, enabling large focal plane areas.  The device used in this work has a 4K$\times$2K format, as used on the Europa Clipper mission. Ultimately for large FOV astronomical applications such as UVEX a 4K$\times$4K size is attractive. Such devices have been made successfully for Lawrence Livermore National Laboratory's National Ignition Facility [\citenum{10.1117/12.2189941}].   All sizes have very similar electrical characteristics since they are fabricated using the same reticles, at the same foundry, using the same processes.
\newline

\begin{table*}[h!]
\begin{center}
{\small
\begin{tabular}{|c|c|} \hline
Total Ionizing Dose & $> 100 $ Krad \\ \hline
Non-Ionizing Energy Loss & $> 3 \times 10^{11} $ protons/$\mathrm{cm^2}$ @ $63$ MeV protons \\ \hline
Single Event Latchup & $> 100$ MeV $\mathrm{cm^2/mg} $ \\ \hline
Single Event Upset & $> 25$ MeV $\mathrm{cm^2/mg} $\\ \hline
\hline
\end{tabular}  
\caption{Radiation tolerance of SRI Nk$\times$Mk, reproduced from data sheet.} 
\label{tab:RadiationTolerance}
}
\end{center}\vspace{-0.5cm}
\end{table*}

\section{Provenance of Sensor Under Test}
\label{sec:Provenance}

The 4K$\times$2K device used for this study was produced together with identical wafers for the Europa Clipper mission [\citenum{10.1117/12.2189941}]. 
The Jet Propulsion Laboratory’s Advanced Detectors, Systems, and Nanoscience Group at JPL’s Microdevices Laboratory (MDL) applied their end-to-end post fabrication processing including bonding, thinning (to $ 7\pm 0.3 \mathrm{\mu}\mathrm{m}$), and delta-doping process using Molecular Beam Epitaxy [\citenum{Burrows10.1142/9446}] [\citenum{hoenk1992growth}] [\citenum{nikzad2017high}] [\citenum{janesick2001scientific}] at wafer level. This process has  been demonstrated over the years on a number of CCD and CMOS formats to provide high and stable quantum efficiency across the UV and visible spectrum [\citenum{nikzad2017high}]. Every photon not reflected or absorbed in the anti-refection (AR) coating is both detected and collected into pixels [\citenum{hoenk1992growth}].

JPL applied various AR coatings and metal dielectric filters (MDFs), returning several die to SRI for packaging and screening tests. The in-band UV QE and out-of-band rejection obtained with the delta doping and direct deposition of MDFs will be described in a companion paper. An earlier broadband AR-coated version was measured for Europa-Clipper band [\citenum{nikzad2021}]

Unfortunately, pinhole defects in the photo-resist (that protects the device during the etching of back surface to reveal the bond pads) caused damage at random locations on the delta doped wafer.  An engineering grade device without any coatings, but with tens of these "etch-through" defects was provided to Caltech for evaluation at lower temperatures than previously explored. At these temperatures several strong glow sources became apparent. 

We will show that the relatively simple on-board circuitry for row/column decoding (no timing generator or ADCs) allows for sufficient glow suppression to meet the demanding dark current requirement with ample margin (\S \ref{sec:darkcurrent}),  while read noise rivals the best CCDs (\S \ref{sec:ReadNoise}).  Linearity is good (\S \ref{Sec:Linearity}), inter-pixel correlation is low(\S \ref{sec:IPC}), and dynamic range is exceptional when CMOS capabilities for dual gain (\S \ref{sec:DualGain}) and dual exposure duration are harnessed. Readout overheads are insignificant. 

\section{SRI Mk$\times$Nk CMOS detector}
\label{sec:SRI}
\subsection{Pixel Topology} \label{pixeltopology}

\begin{figure}[h]
    \begin{center}
       \includegraphics[width=100mm]{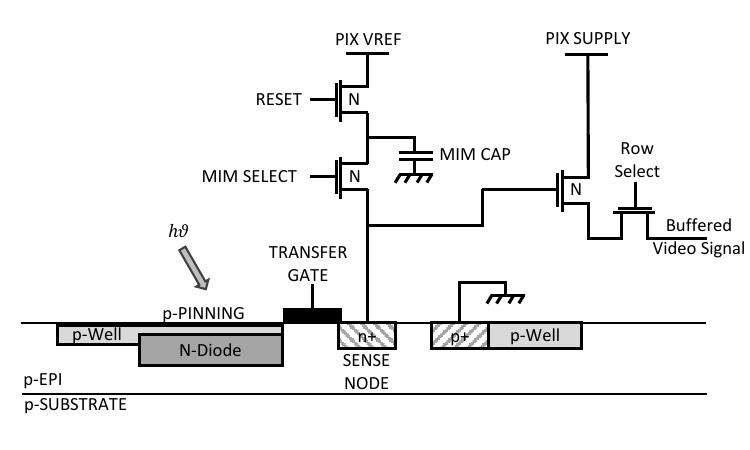}
    \end{center}
\caption{ Five Transistors with Pinned Photo-Diode Pixel (5TPPD) Diagram. Reproduced from [\citenum{10.1117/12.2063524}].} 
\label{Fig:PIXEL}
\end{figure}

Each pixel is made up of a light-sensitive pinned photodiode, a transfer gate, and 4 transistors to reset the sense node and select it during readout (Figure \ref{Fig:PIXEL}). A pixel read cycle begins by resetting the sense node to the reference potential, $\mathrm{PIX\ VREF}$ by closing transistors $\mathrm{RESET}$ and $\mathrm{MIM SELECT}$. If $\mathrm{MIM SELECT}$ is high during the time charge is sensed, the $\mathrm{MIM}$ capacitance is in parallel with the sense node increasing the conversion gain ($\elec/\mathrm{\mu}\mathrm{V}$) and well capacity at the expense of increased read noise (in $\elec$).  By reading out the voltage before and after connecting the MIM capacitor, one can read the same signal in high then low gain, to provide low noise then high signal capacity.  Charge transfer and reset operations are performed for an entire row and cannot be executed on single pixels. This prevents CCD-style Correlated Double Sample (CDS) timing if using off-chip digital subtraction. To support CDS with a conventional short delay between baseline and signal samples, sample-and-hold circuitry is provided at end of each column.  This allows analog subtraction of an entire row to be performed concurrently when reading out in rolling reset mode.

The detector has the ability to reset and transfer charges from the photodiodes to their sense nodes over the whole array concurrently, when a signal labelled $\mathrm{SNAP}$ is pulsed. This  ``Global Shutter" mode results in concurrent exposure start and end for all of the pixels of the array. 

\subsection{Output Buffer} \label{sec:outputbuffer}
The RAW video outputs are buffered by p-type transistors in a Source-Follower topology. We use $3.3\,\mathrm{k\Omega}$ pull-up resistors to $3.3\,\mathrm{V}$ as the load.  No other processing occurs:  we do not invoke the optional RC filtering.

    Alternatively, the CDS video output pins of the detector can be buffered by n-type transistors, also in a Source-Follower topology. In this case an analog CDS circuit located on the bottom of every column, stores the reset levels of a row before subtracting the signal obtained after charge transfer. This subtraction is done using switched capacitors resulting in a 4 $\elec{}$ noise floor (on comparable pixel designs [\citenum{10.1117/12.2063524}]). We prefer Line-wise Digital CDS  since it delivers several times lower noise floor ( \S \ref{sec:ReadNoise})   .

\section{Experimental Setup}
\label{sec:setup}

\subsection{Cryostat} \label{sec:cmost}

A cryostat with Liquid Nitrogen (LN2) cooling is configured for operation of the detector at temperatures down to 130\,K.  A large printed circuit board, dubbed the Vacuum Interface Board (VIB), is located between the front and back sections, sealed with O-rings on each side. The VIB transports all  required signals into the vacuum on internal layers as well as supporting  electrical components in vacuum (such as the zero-insertion-force socket housing the detector) and in atmosphere (FPGA, connectors to detector controller, test points, and switches that support trouble-shooting).
The detector socket is thermally strapped to the nitrogen tank at 77\,K, while the VIB has cut-outs to minimize conduction to the outside world at room temperature. We tuned the thermal link to the liquid nitrogen tank for  120\,K equilibrium temperature without active heating. Detector socket temperature is servo controlled to 140\,K with  millikelvin stability.
Detector electronics are mounted directly on the cryostat in close proximity to the VIB  to limit the length of the video path.

\begin{figure}[h]
    \begin{center}
       \includegraphics[width=150mm]{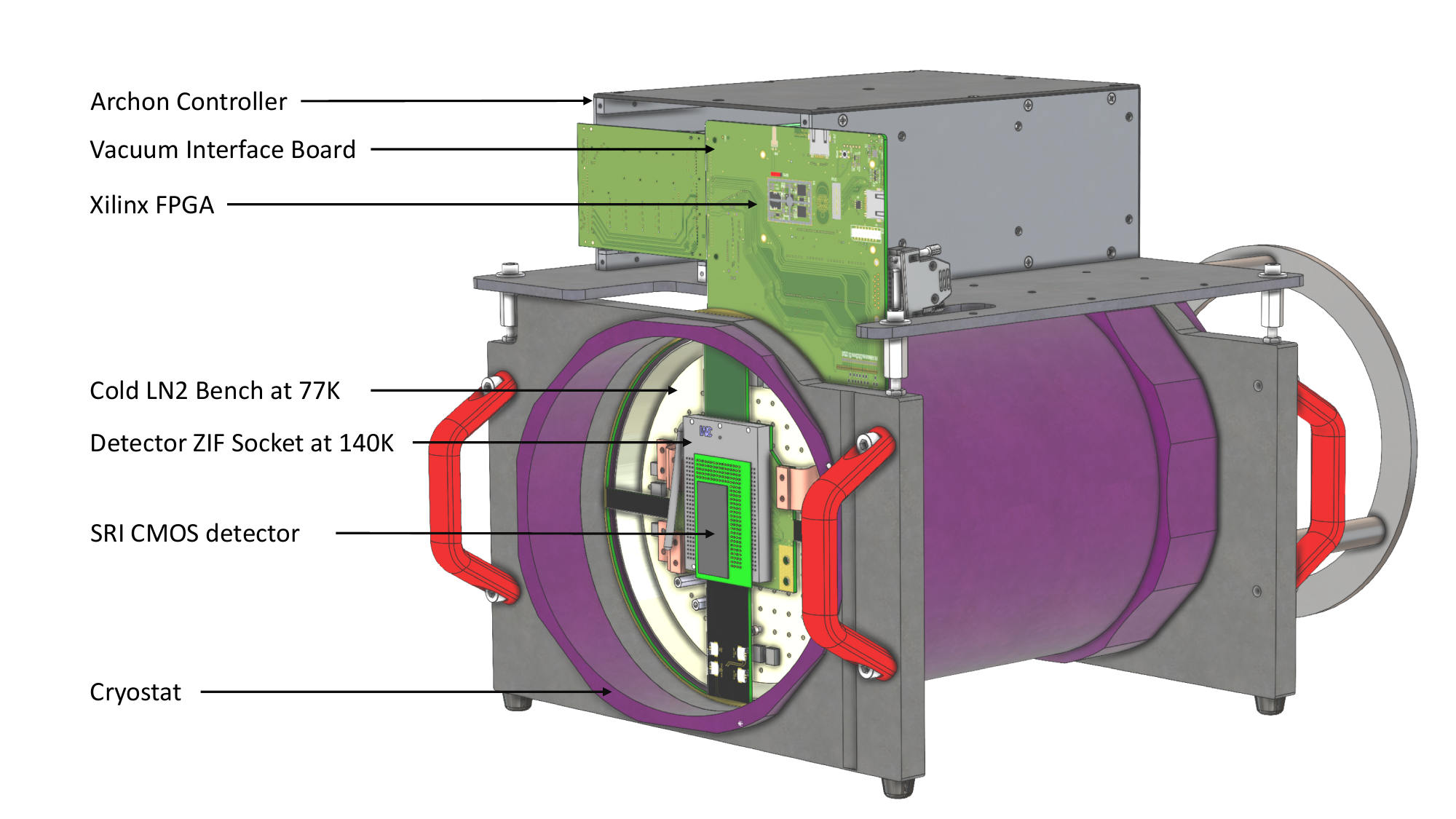}
    \end{center}
\caption{ CMOST Camera. The open cryostat is shown with detector inserted into its zero-insertion-force socket. The controller sits on top of the cryostat, and the Vacuum Interface Board provides row and column address generation and routing of video signals.} \label{fig:cmost}
\end{figure}

\subsection{Camera Controller} \label{sec:cameracontroller}
The detector controller is an ``Archon" made by Semiconductor Technology Associates Inc\footnote{http://www.sta-inc.net/archon/}. It offers versatile configuration including clock and bias drivers, LVDS outputs, and 4 slots, each with 4 video processing channels.  An intuitive waveform definition language has been added by Caltech Optical Observatories (COO).

Direct addressing of rows and columns is supported by the detector to allow efficient subarray readout or user selectable raster scanning direction.  This requires a mix of direct selection of the 1K pixel blocks and binary coding for row selection within those blocks.  As this complexity is not supported by the Archon controller, we have added an external FPGA to generate the addresses, a Xilinx Zynq SoC on Picozed board which plugs into the atmospheric side of the Vacuum Interface Board.  

\subsection{Readout Mode}
The detector offers a large range of possible readout schemes thanks to its versatile addressing topology [\citenum{10.1117/12.2063524}]. In True Global Shutter mode, one must read out a whole frame after Global Reset to acquire baseline pixels and then digitally subtract a newly acquired image after Global Charge Transfer. This frame-wise digital CDS suffers from the long time interval between frame acquisitions, and thus is susceptible to low frequency noise (offset drift). An alternative is to record a single image then subtract the average of many dark frames to remove offset patterns.  While the averaged reference frame is less noisy, it fails to remove $\sim12 \elec$ of $kTC$ noise. To avoid these additional noise sources, all performance measurements reported here have used Rolling Shutter (line by line reset) and line-wise digital CDS.

Rolling Reset is widely used in CMOS detectors.  Light is accumulating at all times, so the transfer of charge from photodiodes to their sense nodes, defines both the end of one exposure and start of the next. Each line is addressed sequentially.  The sense nodes for all pixels in one line are reset concurrently.  In our "line-wise digital CDS" scheme,  baseline samples are digitized sequentially for the entire line, then charge is transferred from photodiodes to all sense nodes concurrently for the same line.  The resulting signal levels for each pixel in the line are then digitized sequentially.   The process repeats for the next line.  The difference between the baseline and signal samples is calculated in the host computer post-facto.  1.5 s is required to scan  a frame with 2048 lines at 700 kpix/s, with 16 channels being read in parallel each servicing a block of 256 columns.  

\begin{figure}[h]
    \begin{center}
       \includegraphics[width=165mm]{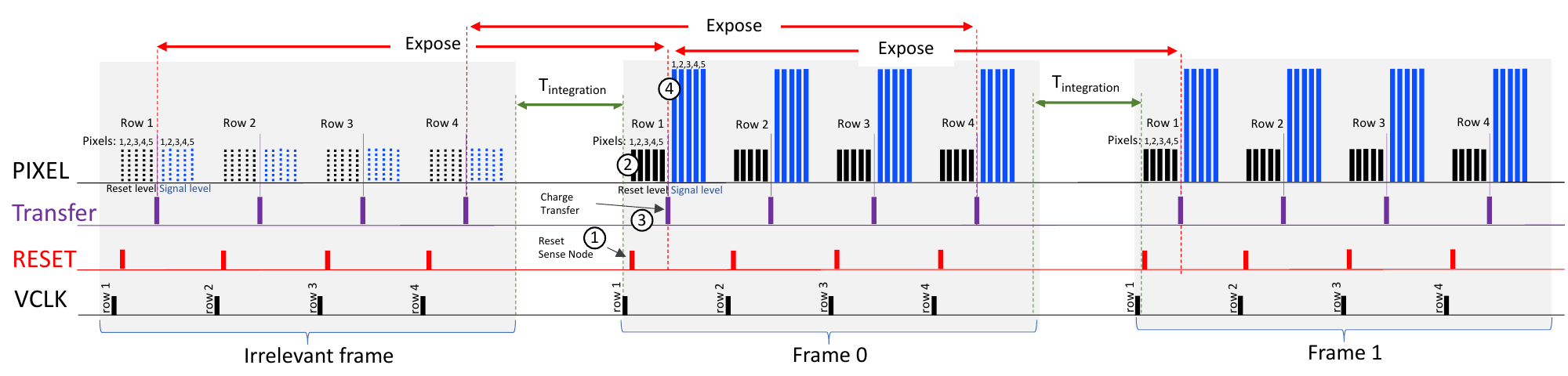}
    \end{center}
\caption{ Readout Diagram of the Rolling Reset Mode. Hypothetical 4 ×4 pixel array. The true exposure time comprises the readout time that is fixed and a parametric exposure time $T_{integration}$ . VCLK is the Vertical Clock pulse that increments the Row Pointer to the next row.}
\label{Fig:DiagramClassicRollingReset}
\end{figure}

The 1.5 s frame time results in a small skew in exposure start times across the image area but this has negligible effect in the long exposures anticipated in our application.  When acquiring a sequence of frames there is a 1.5 s overhead to reset the pixels at start of sequence (e.g. after telescope slew) but no dead time between frames within a sequence.  Table~\ref{tab:ClassicRollingResetSequence} and Figure~\ref{Fig:DiagramClassicRollingReset} describe a multi-frame sequence for a hypothetical $4\,\times4\,\mathrm{pixel}$ array. 

With exposure duration being defined by the time between charge transfers, it follows that minimum exposure time is one frame time and longer exposures are executed by inserting a delay between frame scans. Faster cadence exposures can be executed when reading a subset of rows usually scanning between two preset pointers.

The Archon version used for these experiments was originally developed for CCD readout and thus has an AC-coupled differential video input.  The negative input is grounded while positive input is connected to either the RAW or the CDS output of the detector selected by a mechanical switch.  (We prefer RAW, as noted.)  The DC level of the JFET input stage (in the Archon, post AC coupler) is set through an analog switch typically once per row. This level is chosen so that the differential input voltage is near the beginning of the ADC input range. 
 
Each channel is digitized by a separate 16 bit ADC running at 100\,MHz. An analog filter limits bandwidth into the ADC only to the extent required to prevent aliasing.  The high sample rate allows each channel to be operated in digital oscilloscope mode for diagnostic purposes.   When executing digital CDS, noise bandwidth is set by the number of consecutive 100\,MHz samples which are averaged per pixel.  This averaging occurs concurrently in FPGAs on each video card. 

The line-wise digital CDS arrangement avoids the $kTC$ noise incurred when resetting the analog CDS capacitors.  The fact that baseline and signal have identical signal path assures good gain matching for optimal rejection of low frequencies.  Surprisingly, the 366 $\mathrm{\mu}$s interval between baseline and signal sample proves to be short enough that low frequency noise sources (bias voltage drift; 1/f noise) do not negate these benefits.

\begin{table*}
\begin{center}
{\small
\begin{tabular}{|c|c|c|} \hline
{\bf Step} & {\bf Action} & {\bf Output} \\ \hline\hline
1 & Reset Sense Node & \\ \hline
2 & Read Sense Node & Baseline Image Low Gain\\ \hline
3 & Transfer charge from Photodiode to Sense Node &\\ \hline
4 & Read Sense Node & Signal Image High Gain\\ \hline
\hline
\end{tabular}  
\caption{ Row readout sequence in Rolling Reset mode.Each line is read twice: immediately after reset to capture the baseline and after charge transfer to capture the signal. Timing is shown in Figure \ref{Fig:DiagramClassicRollingReset} }
\label{tab:ClassicRollingResetSequence}}
\end{center}\vspace{-0.5cm}
\end{table*}

\subsubsection{Dual Gain mode}
\label{sec:DualGain}
For best noise performance the smallest possible sense-node capacitance is desired as this increases the voltage change in response to a given charge.  However the signal range is then limited by available voltage swing.  To allow greater signal range, a MIM capacitor in each pixel can be added in parallel to the sense node capacitance by closing the appropriate analog switch (Figure \ref{Fig:PIXEL}).  Since the readout of the voltage does not affect charge on the sense node, it is possible to read the same charge packet first with MIM capacitor isolated then again after it is connected. The sequence of steps for this Dual Gain readout mode is detailed in Table~\ref{tab:HDRSequence}. 

The key here is to establish separate CDS baselines for both gain states.  There is no $kTC$ noise problem since any charge trapped when opening the gain-select switch is measured at Step 4 and vanishes when the switch is closed again at Step 7.

Frame time is almost doubled since each line is digitized twice as many times.  Due to the added time delay, the difference between the frames produced at Steps 9 and 2 provides a CDS sample which is inferior to the difference between frames produced at Steps 6 and 4. However, this slightly elevated read noise is irrelevant since the low-gain sample is only used for large signals that are shot-noise limited.   

\begin{table*}
\begin{center}
{\small
\begin{tabular}{|c|c|c|} \hline
{\bf Step} & {\bf Action} & {\bf Output} \\ \hline\hline
1 & Reset Sense Node in Low Gain & \\ \hline
2 & Read Sense Node in Low Gain & Baseline Image Low Gain\\ \hline
3 & switch to High Gain &\\ \hline
4 & Read Sense Node in High Gain & Baseline Image High Gain \\ \hline
5 & Transfer charge from Photodiode to Sense Node &\\ \hline
6 & Read Sense Node in High Gain & Signal Image High Gain\\ \hline
7 & Switch to Low Gain & \\ \hline
8 & Transfer charge left behind from Photodiode to Sense Node. \textbf{KEY STEP} & \\ \hline
9 & Read Sense Node in Low Gain & Signal Image Low Gain\\ \hline
\hline
\end{tabular}  
\caption{ Row readout sequence to generate a dual-gain image. Each row is read four times to generate two pairs of row per gain. A second charge transfer must be performed after switching to low gain to recover the full signal range.}
\label{tab:HDRSequence}
}
\end{center}\vspace{-0.5cm}
\end{table*}

\section{Performance}
\label{sec:results}

\subsection{Photon Transfer Curves and Well capacity}
To infer well capacity and conversion gain $(\elec/\mathrm{ADU})$, pairs of identical exposures with increasing integration time are acquired with temporally stable flux and moderate spatial uniformity. We then subtract the two frames of each pair to remove common mode patterns due to illumination or fixed electrical patterns, leaving only the random noise component of the signal.  Spatial variance is then calculated for a region chosen to be free of bad pixels or other artifacts such as ROIC glow. This procedure is repeated for both low and high gains.

\begin{figure}[h]
    \begin{center}
       \includegraphics[width=165mm]{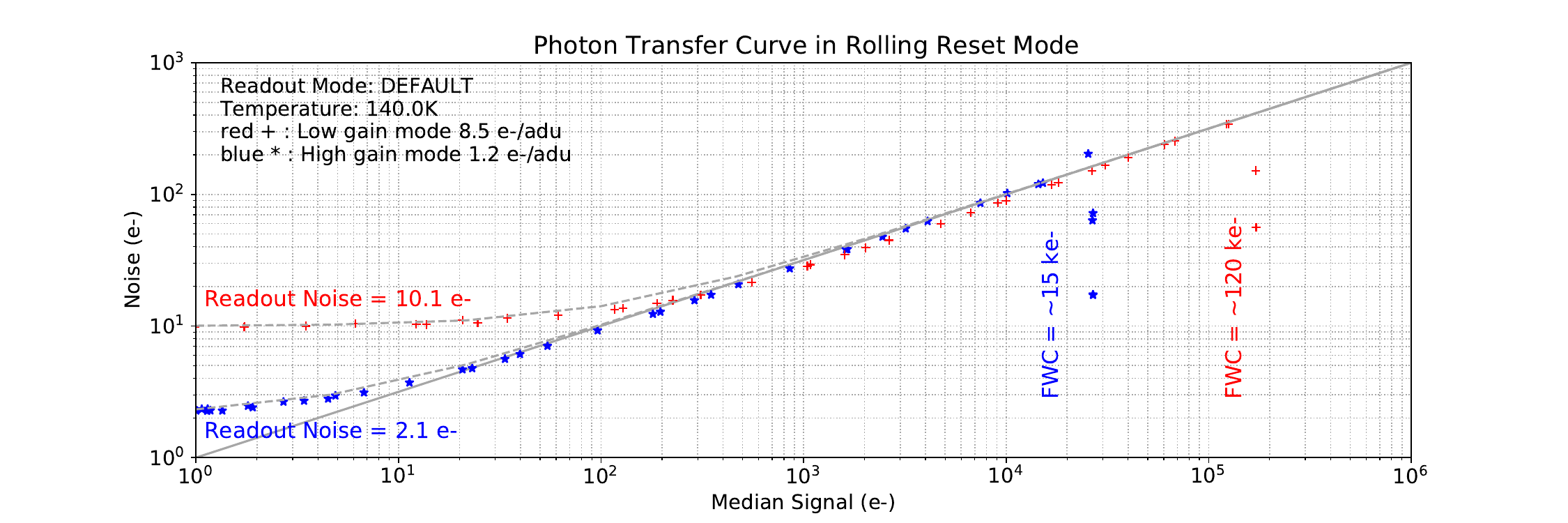}
    \end{center}
\caption{ Photon transfer curves of low and high gain modes. Readout noise, system gain and Full Well Capacity can be directly read from this plot.} 
\label{fig:transfercurve}
\end{figure}

Conversion Gain ($\elec/\mathrm{ADU}$), the inverse of the slope of the the linear fit, is calculated separately for each gain mode. These values are then used to convert signal and variance to values displayed in electrons, for both gain states on the same Photon Transfer Curve (Figure~\ref{fig:transfercurve}).
This plot format is convenient to immediately highlight the signal level at which shot noise dominates.

When only spatial variance is used in photon transfer curves, total noise is underestimated due to interpixel correlations, so system conversion gain $(\elec/\mathrm{ADU})$ is over estimated. The correct conversion gain is obtained when the covariance with adjacent pixels is included in the photon transfer curve . Such interpixel correlations were observed only in low-gain mode (\S \ref{sec:IPC}). The 8.0 $\elec/\mathrm{ADU}$ conversion gain in low-gain mode has been corrected for Inter-Pixel Correlation (IPC) throughout this paper resulting in a 6\% reduction in dark current and read noise compared to that inferred from simple photon transfer method using variance alone.  The photon transfer curve then yields readout noise of 2.1\,$\elec$ and 9.5\,$\elec$ and a Full Well Capacity of 15\,$\mathrm{k}\elec$ and 113\,$\mathrm{k}\elec$ in the high gain and low gain modes, respectively.

\subsection{Linearity}
\label{Sec:Linearity}
Linearity has been measured in both high and low gain modes by varying the exposure time while the detector is illuminated with a constant flux. To do this, a red LED is attached to the LN2 tank on a separate thermally regulated platform, operating slightly above 77\,K. A bench current source injects a current ranging from a few tens to hundreds of nanoamperes leading to a total power dissipation well below a microwatt, resulting in very little efficiency variation due to junction self-heating.   

We infer that detector self-heating is responsible for some gain drift as readout cadence influences the level of the video signal. Sensor current consumption peaks at $\sim70\,\mathrm{mA}$ during readout, dissipating hundreds of milliwatts. This highlights the need for a more thermally conductive package to minimize the change in detector temperature due to self heating.  The  package in use is far from optimal in this regard so significant performance improvement is probably possible. It is also clearly beneficial to adopt clocking schemes which maintain the readout cadence even when idle to minimize temperature drift.

\begin{figure}[h]
    \begin{center}
       \includegraphics[width=165mm]{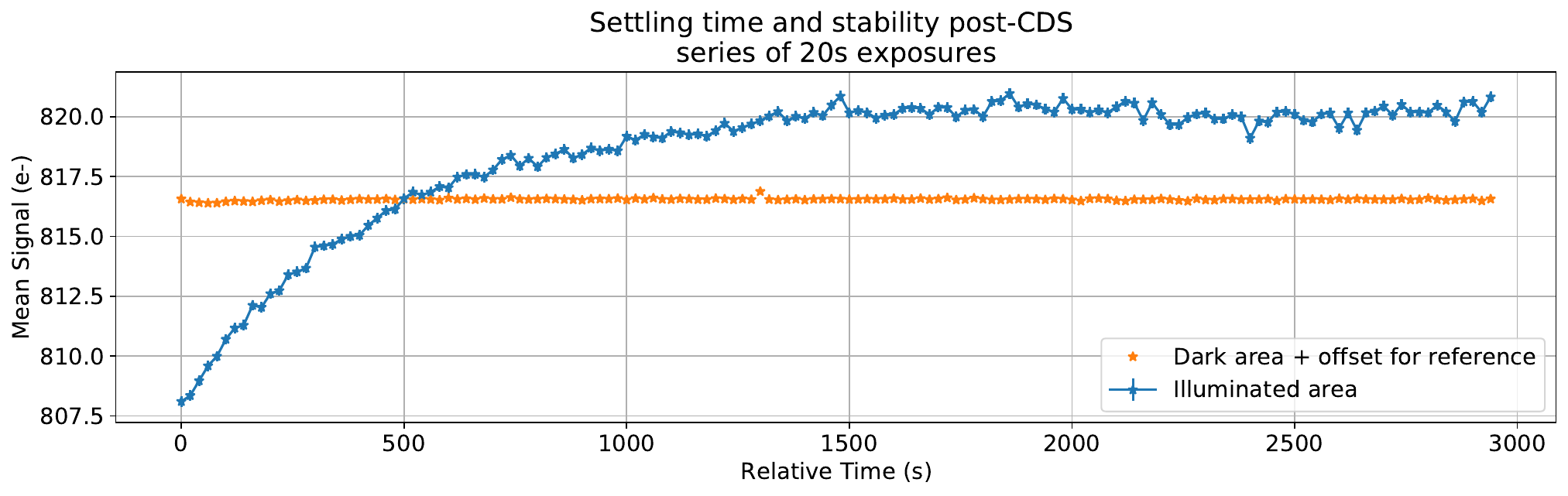}
    \end{center}
\caption{Initial transient after starting the readout of the detector at a constant (and fastest) frame rate. Each data point is the average of a clean $128\times128$ window. The transient is only visible in the illuminated area, implying that this is a gain change and not an offset.} 
\label{Fig:TransientLevel}
\end{figure}

An initial transient in the signal is seen in Figure~\ref{Fig:TransientLevel} after turning on the detector and continuously reading it at a constant cadence. This transient appears only where there is illumination: not in dark areas under a mask. This demonstrates that the CDS processing indeed corrects for offset drift of the RAW video levels but cannot prevent the gain change. We explore this dependence of gain on temperature in more detail in \S \ref{sec:GainTempCo},
finding that the video level settles to  1$\elec$ after approximately 30\,min. 

In order to mitigate gain changes due to self-heating during the linearity experiment, the detector was read out at a very low duty cycle (at least 5 min per frame), so that the effect of self-heating was negligible.

We estimated the non-linearity by fitting an affine function or a quadratic with non-zero value at origin. We acknowledge that this is not ideal, but fixing a zero offset leads to a large mismatch with our fit.
In order to optimize both the residuals during our fitting process and the non-linearity itself, we included a weighting factor, $w_i$, in the estimator $E$, which is defined as

\begin{equation}
    E = \sum w_i^2\left(p(x_i)-y_i\right)^2 = \sum \left(\frac{p(x_i)-y_i}{y_i}\right)^2
\end{equation}

\noindent where $y_i$ is the measured signal at exposure time $x_i$, and $p(x_i)$ is the value of the linear fit at exposure time $x_i$. This prevents the gain correction from increasing steeply at the lower end of the non-linearity curve when the denominator becomes smaller.

\begin{figure}[h]
    \begin{center}
       \includegraphics[width=165mm]{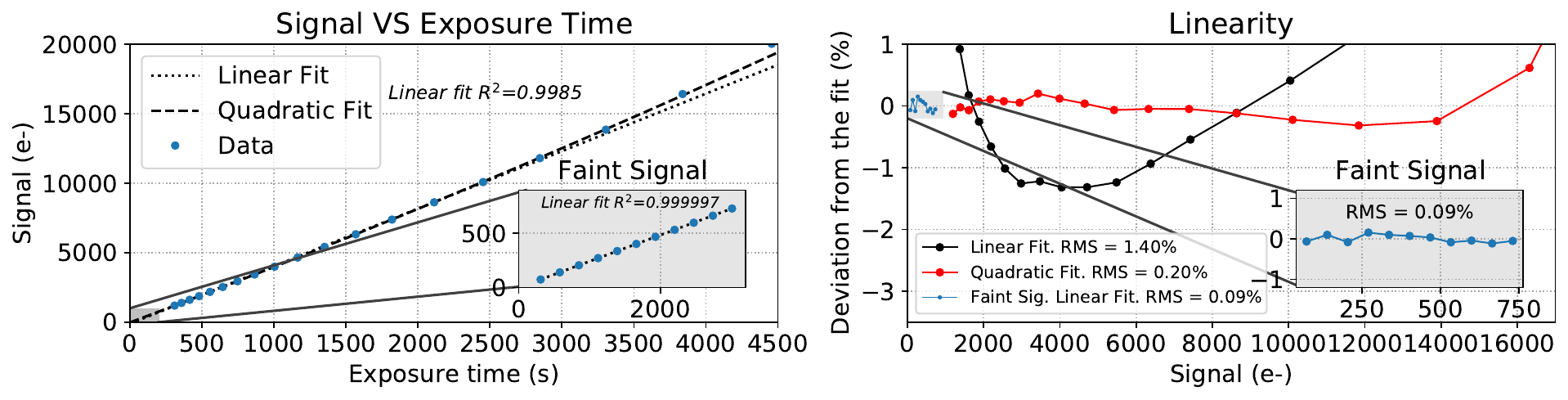}
    \end{center}
\caption{Linearity in high gain mode. Signal is the median of a 256x256 window. Faint signal ranges from 0\,$\elec$ to 1000\,$\elec$. An F-test comparing a quadratic fit to a linear fit returns a p-value = $2.2\times10^-13$, confirming that the quadratic model provides a statistically improved fit.}
\label{Fig:LinearityHighGain}
\end{figure}

Figure \ref{Fig:LinearityHighGain} summarizes the linearity behavior in high gain ($1.2\,\elec/\mathrm{ADU}$) mode. A linear fit results in an RMS deviation of 1.4\%, with larger disagreement for higher signal. The residuals show clear curvature, motivating a quadratic fit to the data. Better agreement is achieved with a quadratic fit, yielding an RMS of 0.2\% and flatter residuals. We perform an F-test and find that the quadratic fit is statistically required, with the null hypothesis of a linear fit rejected with $p = 2.2\times10^{-13}$. This is consistent with typical CMOS sensors where non-linearity originates in both the voltage-dependency of the PN junction capacitance that contributes to Sense Node capacitance, and in the non-linearities of the source-follower itself [\citenum{89d8024d8ae24418a1405aa73ec33d21}]. The inset plot shows linearity behavior for faint signals (obtained at reduced flux), demonstrating that the fit improves to an RMS deviation of 0.1\% and projects to 0 signal for null exposure time. (We must extrapolate to zero-length exposure since minimum exposure time is the frame scan time which is 1.5\,s when reading at $700\,\mathrm{kPix/s/ch}$.)

\begin{figure}[h]
    \begin{center}
       \includegraphics[width=165mm]{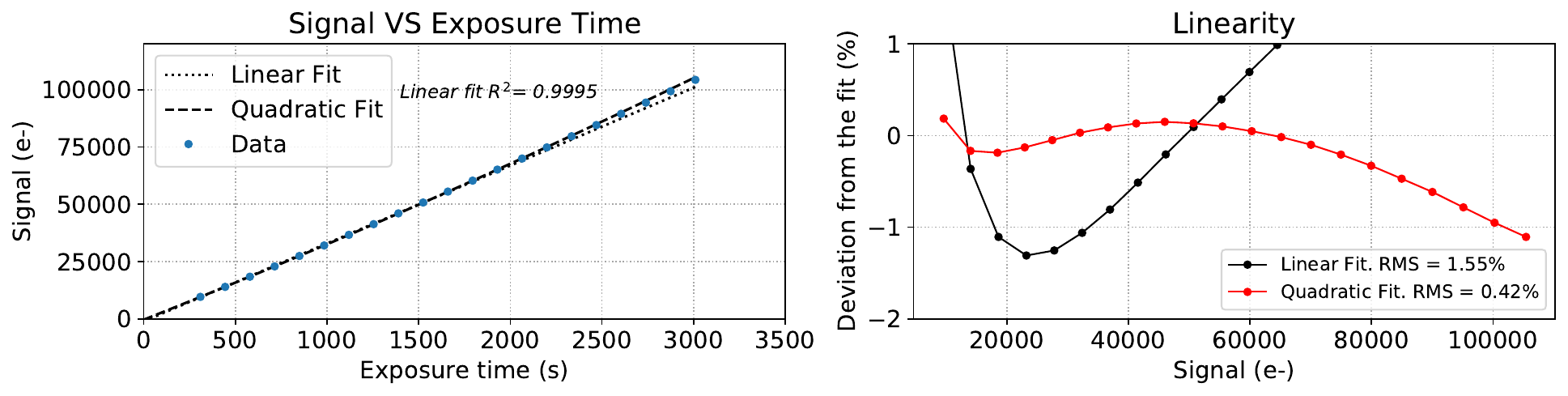}
    \end{center}
\caption{Linearity in low gain mode. Signal is the median of a 256x256 window. Note that the faint signal is neglected since low gain mode will not be utilized for very faint signals. An F-test comparing a quadratic fit to a linear fit returns a p-value = $2.6\times10^-12$, confirming that the quadratic model provides a statistically improved fit.}
\label{Fig:LinearityLowGain}
\end{figure}

Figure \ref{Fig:LinearityLowGain} summarizes the linearity behavior in low gain ($8.0\,\elec/\mathrm{ADU}$) mode. This shows similar behavior: the linear fit yields an RMS value of 1.5\%, while the quadratic fit gives better agreement with RMS deviation 0.4\% and is similarly a statistically-justified improvement over the linear fit. Note that the faint signal is neglected since low gain mode will not be utilized for very faint signals.

\subsection{Temperature dependence of the gain} \label{sec:GainTempCo}
Temperature stability requirements in flight will probably be driven by the temperature dependence of the gain. To evaluate this, we recorded identical flat fields at moderate intensity for temperatures varying between 140\,$\mathrm{K}$ to 180\,$\mathrm{K}$. Light intensity was held constant while regulating LED temperature to millikelvin precision. We allowed a settling time of 30\,min between each 5\,$\mathrm{K}$ step.

\begin{figure}[h]
    \begin{center}
       \includegraphics[width=100mm]{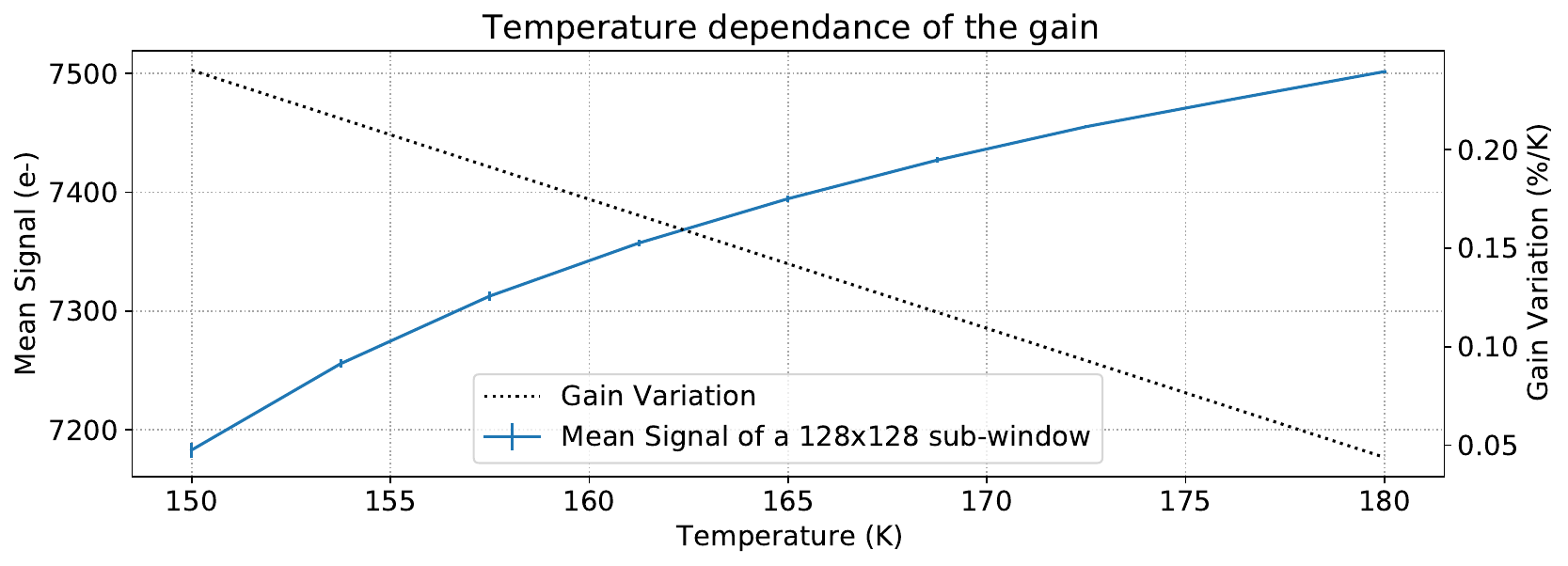}
    \end{center}
\caption{Temperature dependence of the gain between 150\,$\mathrm{K}$ and 180\,$\mathrm{K}$. Around 165\,$\mathrm{K}$, the gain varies by 0.15\,$\%/\mathrm{K}$. The signal is computed as the mean of a 128x128 window.}
\label{Fig:GainTempCo}
\end{figure}

Figure~\ref{Fig:GainTempCo} shows that the temperature dependence of gain is not linear but gradually flattens as the temperature increases. Around 165$\,\mathrm{K}$, the gain varies by 0.15\,$\%/\mathrm{K}$. It is not possible to determine from this experiment whether the gain dependence on temperature is caused by variation in QE, sense node capacitance, Source-Follower transconductance, or load resistance. 

It would be interesting to reproduce this experiment to characterize the temperature dependence of the offset.  While the per-row CDS processing removes any slow DC-offset variation, one could track the baseline samples, however, the AC-coupling of our video acquisition board currently prevents such measurement.

\subsection{Multiplexer (Mux) glow}
\label{sec:MUxGlow}
\begin{figure}[h]
    \begin{center}
       \includegraphics[width=160mm]{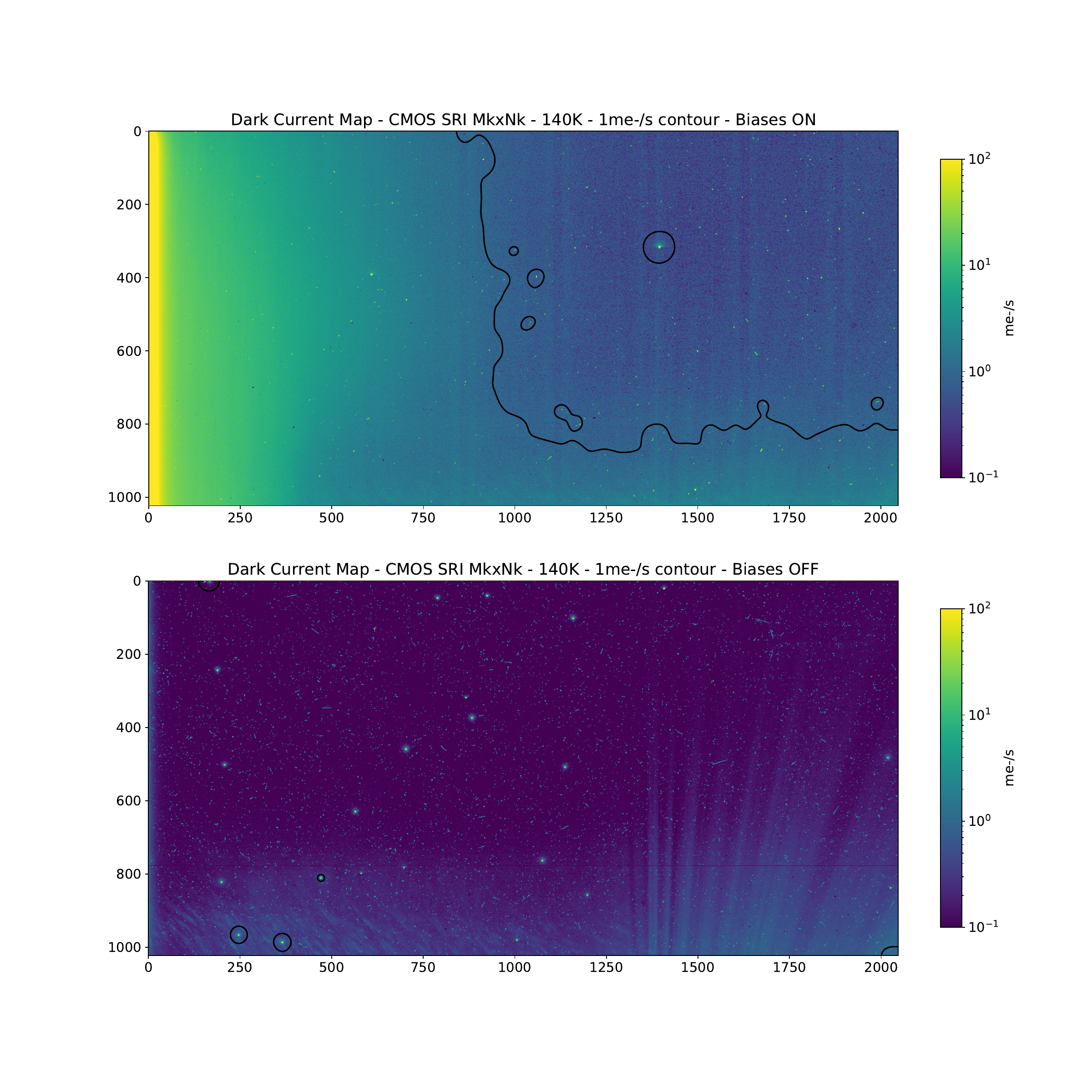}
    \end{center}
\caption{ Top Figure: 18 hours exposure in the dark while keeping electronics in its default state. Bottom Figure: ``Low glow" 18 hours exposure in the dark while shutting down support electronics. \\ Dark Maps of the top left quadrant of the detector show fine structures and glow sources in and around the image area. Most of the glow coming from the left edge of the chip, where the logic and pixel drivers are located, is turned off by lowering the biases during the exposure.  Charge collection is unaffected.}  
\label{fig:LongDarkMapBiasONOFF}
\end{figure}

Electronics surrounding the image area (primarily logic and pixel drivers on the left edge of the chip) act as glow sources that become visible in long exposures. Figures~\ref{fig:LongDarkMapBiasONOFF} shows Dark Maps generated  from the difference of 18 hour and 1 hour exposures to suppress fixed patterns and reveal $\mathrm{m\elec/s}$ effects. 

It can be seen that light is emitted by the row select logic to the left of the image and the column select logic well below the bottom of the image. The device under test also had several strong glow sources in the image area to the right and below of the area displayed.

We suppress photo-emission from the address logic by reducing the levels of bias voltages during exposure, where not required in the charge collection process. The only remaining biases are \texttt{VDD} driving the logic and the Transfer Gate that keeps the charges in the photo-diode. Experience shows that the three biases \texttt{V\textunderscore Ref}, \texttt{V\textunderscore Reset\textunderscore High}, and \texttt{V\textunderscore MIM\textunderscore High} must be kept on to ensure a valid bias of the Sense Node before readout. Some settling time was added after turning the other biases on again and starting a readout of the image area. This settling time, though dozens of seconds, is negligible compared to the long integration times required for glow to be visible. This glow-mitigation technique greatly increased the area where dark current is less than  1\,$\mathrm{m\elec/s}$.

\begin{figure}[h]
    \begin{center}
       \includegraphics[width=160mm]{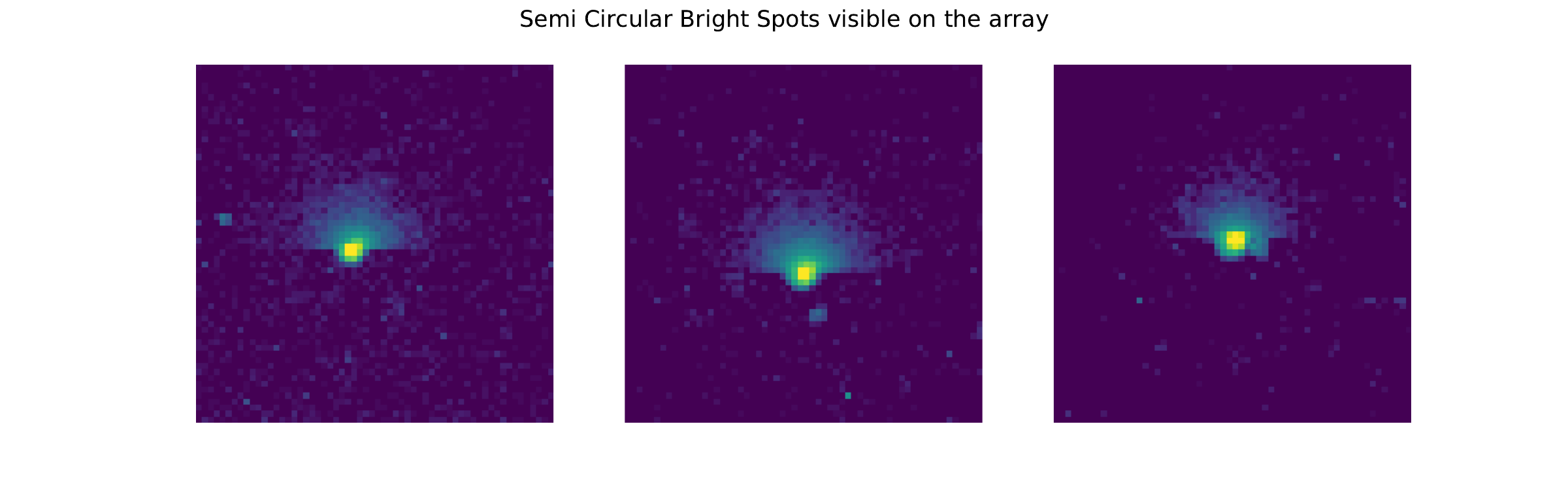}
    \end{center}
\caption{ Some bright spots show a peculiar semicircular shape with the same orientation.}
\label{fig:SemiCircularSpots}
\end{figure}

\subsection{Dark current} \label{sec:darkcurrent}
Dark current has been measured at temperature between 140\,$\mathrm{K}$ and 240\,$\mathrm{K}$. One hour was allowed for settling between each temperature change. No hysteresis from heating or cooling the detector is observed. We evaluate dark current within a sub-window in the region of the detector minimally affected by glow.  In spite of the very low signal,  good signal to noise ratio is obtained by using a large number of pixels and very long exposure times.

McGrath et al. [\citenum{McGrath:2018:2470-1173:354}] list different mechanisms as possible sources of dark current, each having their own temperature dependence. We do not attempt to identify the dominant source,  noting only that the dark current decrease follows the expected Arrhenius law, until a a floor is reached at 0.08\,$\mathrm{m\elec/s}$ for temperatures below 165\,$\mathrm{K}$. 

Given 2\,$\elec$ read noise and 900\,s exposure time, the dark current can rise to 1\,$\mathrm{m\elec/s}$ before increasing total noise by 10\%. These requirements imply that the operating temperature could be as high as 184\,$\mathrm{K}$ before shot noise on dark current becomes significant.

The striations to the right in Figure~\ref{fig:LongDarkMapBiasONOFF} (Bottom "low-glow" Map) are due to light scattering from a bright defect at the lower edge, which we assume will not be present in science grade devices. A dozen or so larger, approximately semicircular  white spots are are not due to cosmic rays but appear to represent weak glow sites of order $2\,\mathrm{m\elec/s/pixel}$ at peak, affecting fewer than 1\% of pixels (Figure \ref{fig:SemiCircularSpots}).

\begin{figure}[h]
    \begin{center}
       \includegraphics[width=160mm]{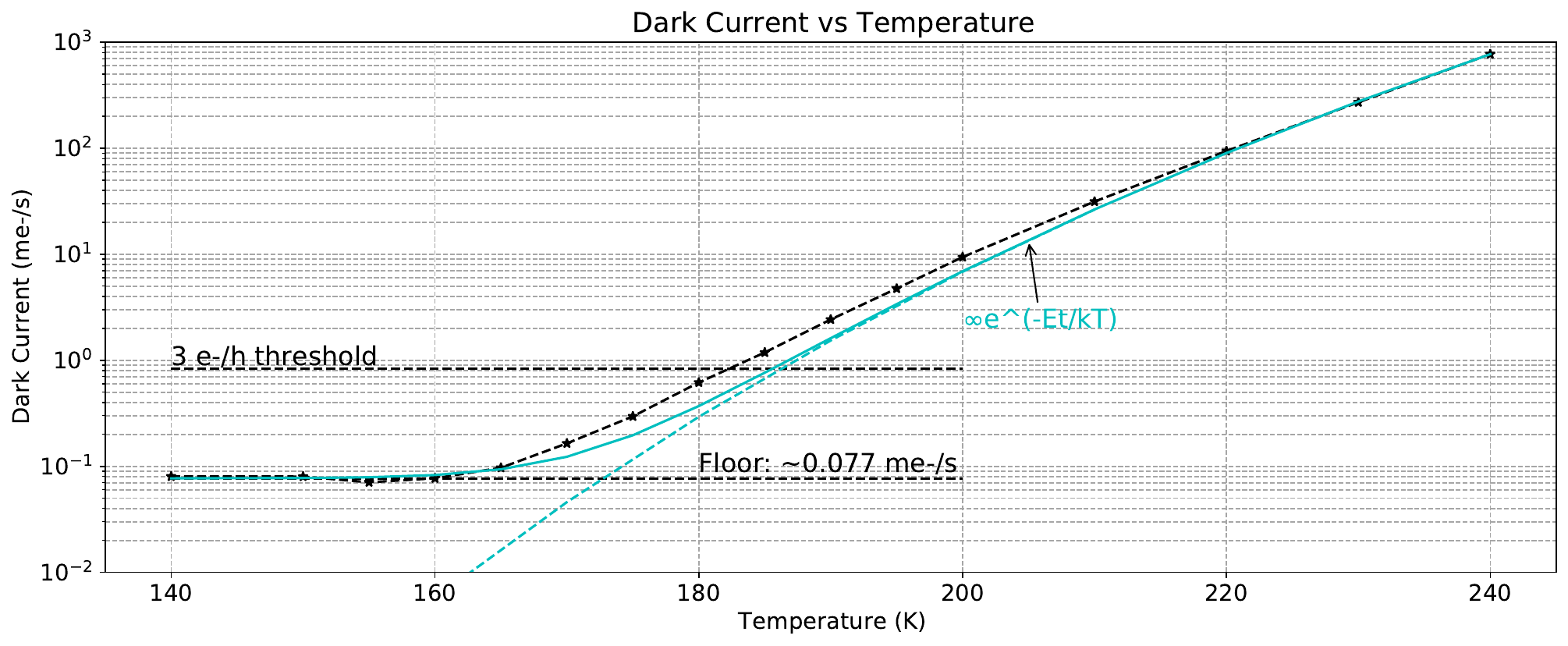}
    \end{center}
\caption{ Dark current for a wide range of temperatures. A simple fit of an Arrhenius law shows the expected temperature dependency.} 
\label{fig:DarkCurrentvsTemp}
\end{figure}

\subsection{Optimizing Speed and Noise Performance}

The STA Archon controller digitizes pixels at a rate of 100\,MHz and supports sample averaging in real time for each channel to provide low-pass filtering. This is the digital equivalent of the analog integrator found in traditional CDS processors.  

Readout noise tends to increase with higher pixel frequency, since noise bandwidth increases when there is less sample averaging to reject high and mid frequency noise. Conversely, when pixel rate is low the increased interval between baseline and signal samples results in poorer rejection of low frequency noise processes which begin to dominate. To find the noise minumum between these extremes, we explored multiple pixel frequencies from 700\,kHz to 1.4\,MHz for numerous averaging window widths.

For any given pixel time, when the averaging window is too wide, the sampling of the video signal begins before it settles sufficiently, with adverse effects on gain and linearity behavior. Therefore we check Gain and linearity using low level flat illumination to infer changes in system gain and thus the true noise in electrons.

\begin{figure}[h]
    \begin{center}
       \includegraphics[width=150mm]{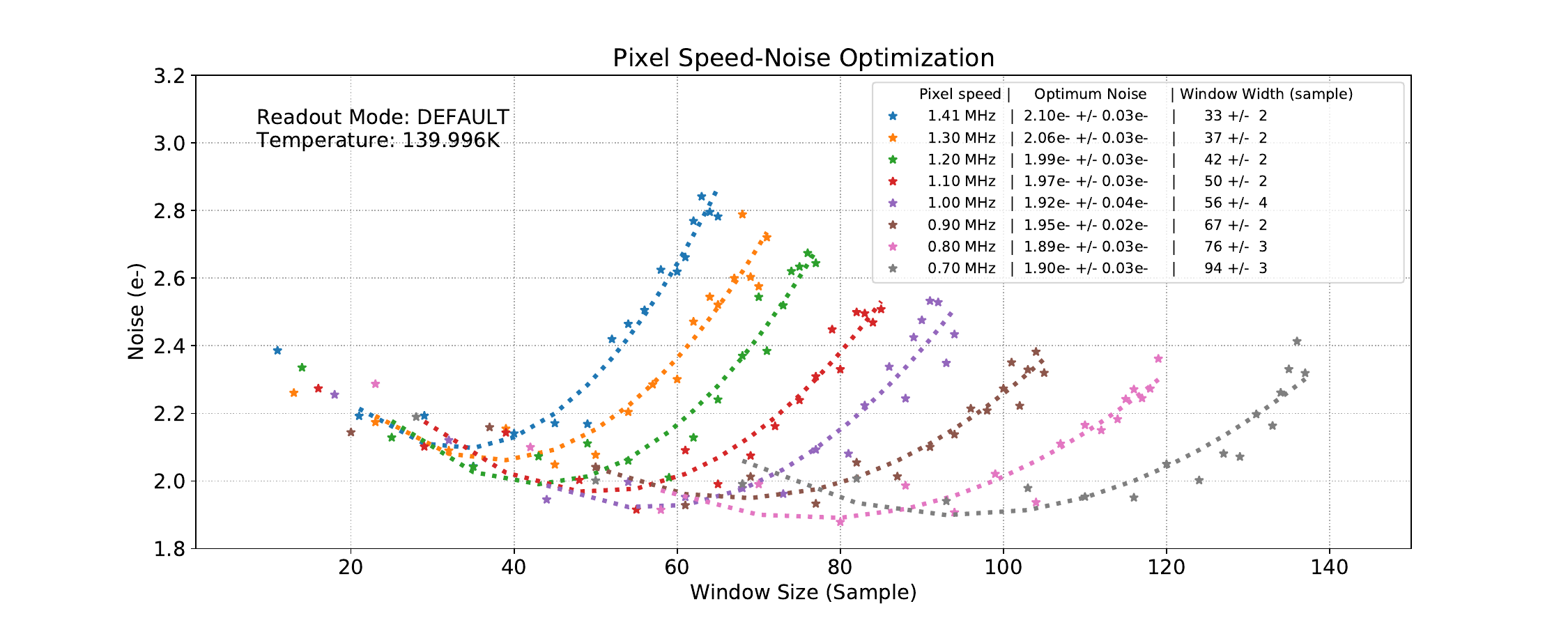}
    \end{center}
\caption{Pixel Speed Noise optimization. The noise is defined as the spatial standard deviation of a 100x100 sub-window of the array. For each pixel frequency, an optimum number of samples is found by fitting a parabola to the data.}
\label{fig:SpeedNoise}
\end{figure}

When the window becomes too wide, the gain decreases and the noise rises. For each pixel frequency, there is therefore an optimal window width at the bottom of each curve in Figure \ref{fig:SpeedNoise}. In this paper, we only use the optimum sample width for any given pixel frequency.

\subsection{Read noise}
\label{sec:ReadNoise}
CMOS sensors can exhibit significant spatial variation in readout noise, predominantly due to single electron traps close to the channel of the pixel buffer MOSFET. On some pixels, this creates a bistable video level known as Random Telegraph Noise (RTN), whose amplitude and characteristic time constant vary from pixel to pixel, with characteristic switching rate being  dependent on temperature.

\begin{figure}[H]
    \begin{center}
       \includegraphics[width=170mm]{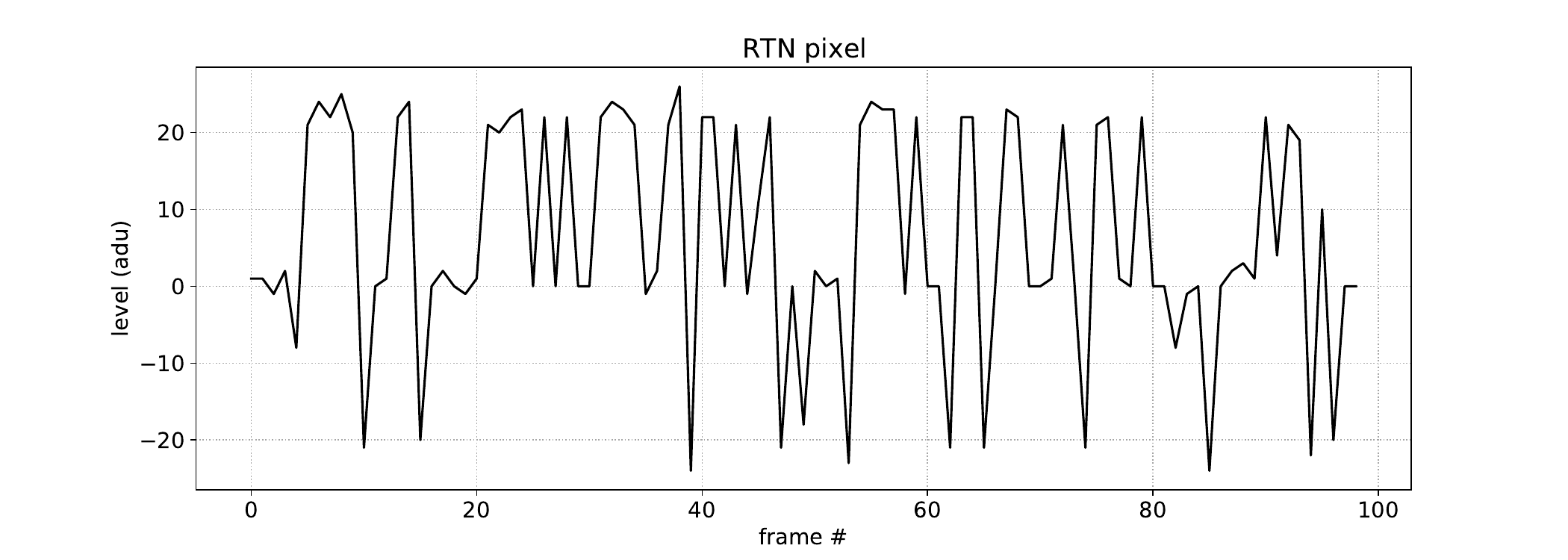}
    \end{center}
\caption{Fluctuating level of a pixel affected by Random Telegraph Noise. The amplitude of the step is much higher than the read out noise. Note that the size of the step is constant while transition timing is random.}
\label{fig:RTNPixel}
\end{figure}

In Figure~\ref{fig:RTNPixel} a time series of a pixel exhibiting strong RTN can be seen jumping between three states after CDS processing, representing a negative transition between CDS samples, a positive transition, or no transition. 

\begin{figure}[h]
    \begin{center}
       \includegraphics[width=170mm]{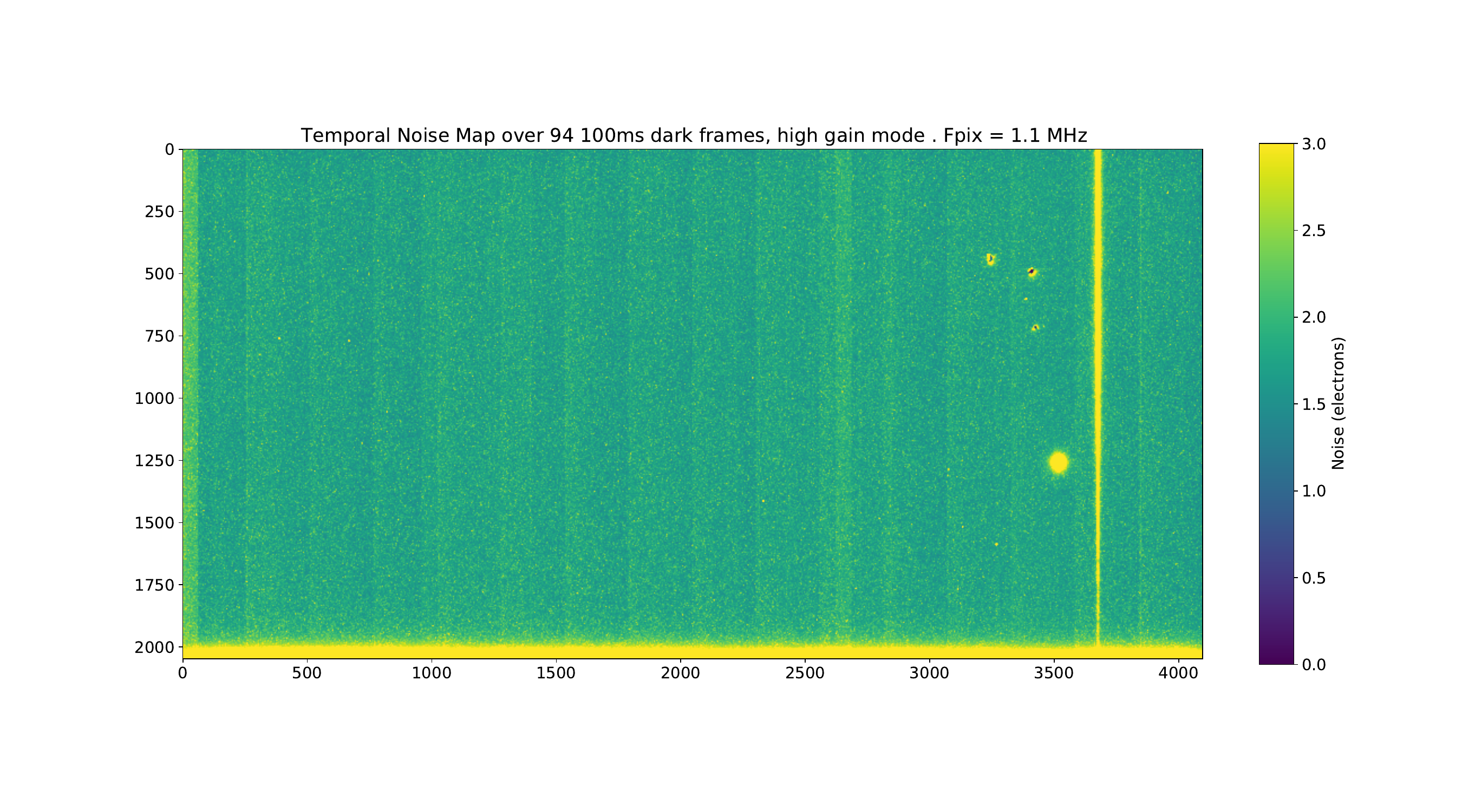}
    \end{center}
\caption{Noise Map of the detector in high gain. Excess noise is visible in the bottom rows, next to the amplifiers. The bad column  generates noise in its surrounding as well as four distinct clusters.}
\label{Fig:NoiseMap}
\end{figure}

The noise map in Figure~\ref{Fig:NoiseMap} was created by computing the (temporal) standard deviation of each pixel from 94 short dark exposures.

\begin{figure}[H]
    \begin{center}
       \includegraphics[width=170mm]{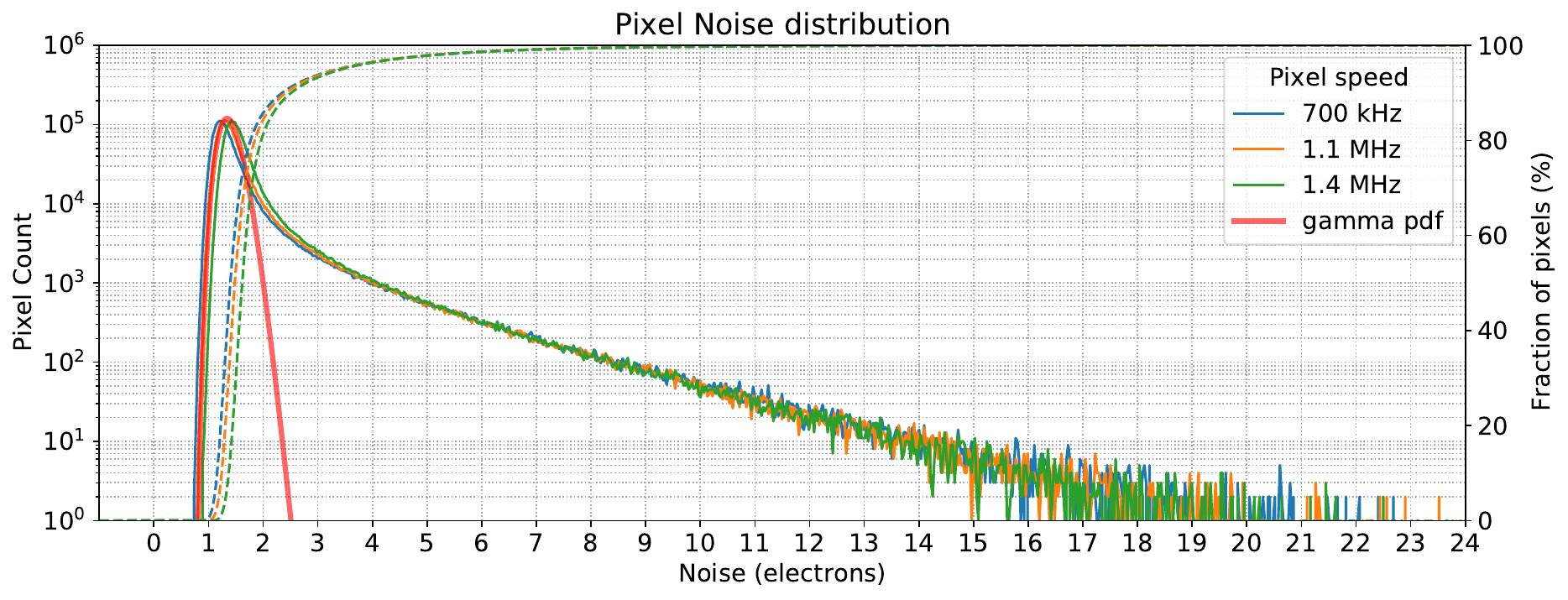}
    \end{center}
\caption{Noise distribution of pixels in a clean sub-array of the detector. The ideal gamma distribution describes distribution if noise were distributed normally in the detector. The dashed curves represent the fraction of pixels (right y-axis) below a given noise level (x-axis) }
\label{fig:NoiseHistogram}
\end{figure}

The histogram in Figure~\ref{fig:NoiseHistogram}, was created from data in subset at top left of the temporal noise map, chosen to exclude defective areas. Instead of following the ideal Gamma distribution, the measured distribution is skewed with many pixels being noisier than expected.

\begin{figure}[H]
    \begin{center}
       \includegraphics[width=140mm]{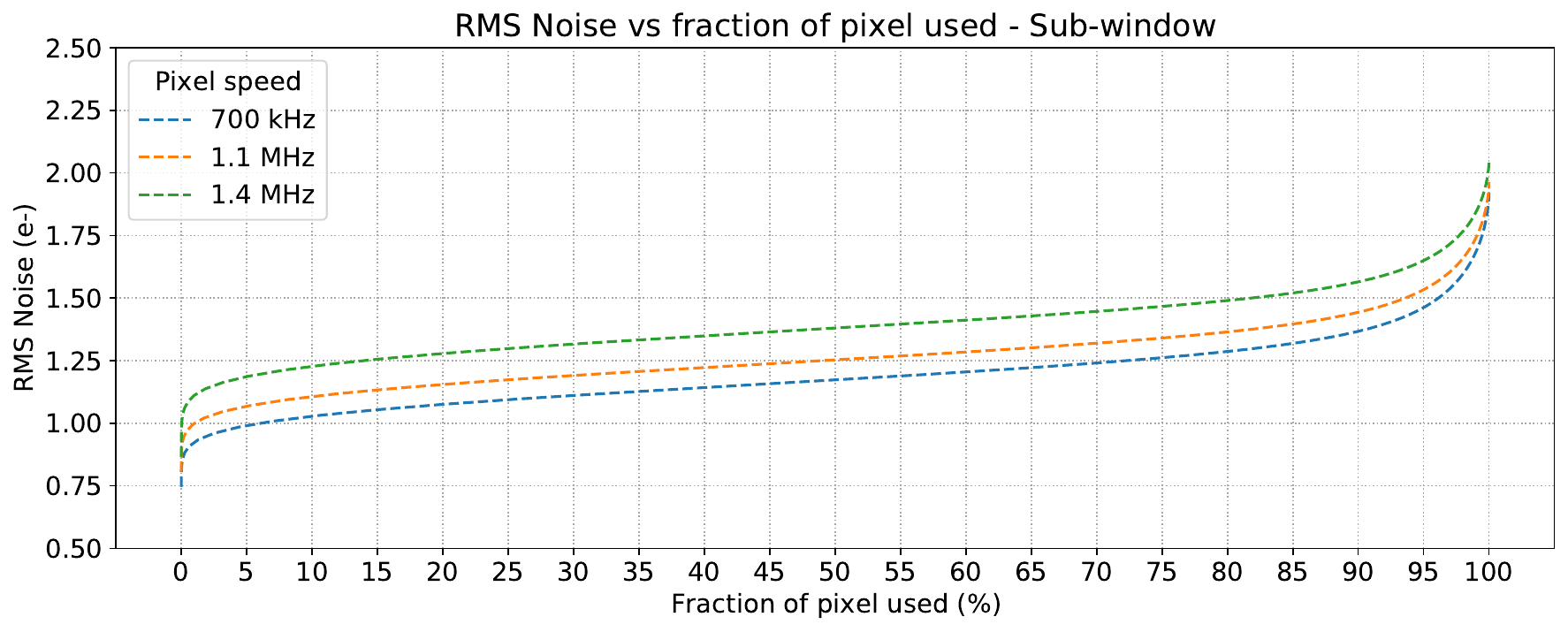}
    \end{center}
\caption{Square root of average noise variance for all pixels included in window, after exclusion of noisiest pixels, plotted for three different pixel rates.}
\label{fig:RMSNoise}
\end{figure}

Given this highly skewed distribution, it may not be obvious what is the best metric for noise performance. The modal noise is 1.33\,$\elec$, the median noise is 1.43\,$\elec$, and the mean is 1.70\,$\elec$. We propose that a better metric is the square root of the average variance across all pixels not excluded as ``too noisy to be used".  The square of this ``RMS Noise" times the number of pixels in a photometric aperture then represents expected total variance which can be used for making SNR projections.  Figure~\ref{fig:RMSNoise} shows the RMS Noise as a function of the number of pixels included.  As more pixels are classified as ``bad" the RMS Noise improves but at the expense of lost information where interpolation across bad pixels is employed.

\subsection{Inter-Pixel Correlation (IPC) - Cross-correlation}
\label{sec:IPC}

Pixels may couple to their neighbors via two main effects [\citenum{Hirata:2019:10.1088/1538-3873/ab44f7}][\citenum{Moore:2004:10.1117/12.507330}]: 1) Brighter Fatter Effect (BFE) [\citenum{arXiv:1402.0725v1}] [\citenum{2018PASP..130f5004P}] is the tendency for previously accumulated charge to repel incoming charge, allowing neighboring pixels to collect additional electrons; 2) Inter-Pixel Capacitance leads to voltage coupling between neighboring pixels [\citenum{Choi:2019:10.1088/1538-3873/ab4504}]. The method described below is used to evaluate the total contribution of inter-pixel effects (BFE and Inter-Pixel Capacitance) to cross-correlations without distinguishing them. This total effect will be referred to as the Inter-Pixel Correlation (IPC). 
A method to distinguish underlying inter-pixel phenomena using correlations in flat fields is described in papers by Hirata et. al. [\citenum{Hirata:2019:10.1088/1538-3873/ab44f7}] and Choi, et. al. [\citenum{Choi:2019:10.1088/1538-3873/ab4504}].

\begin{figure}[h]
    \begin{center}
       \includegraphics[width=190mm]{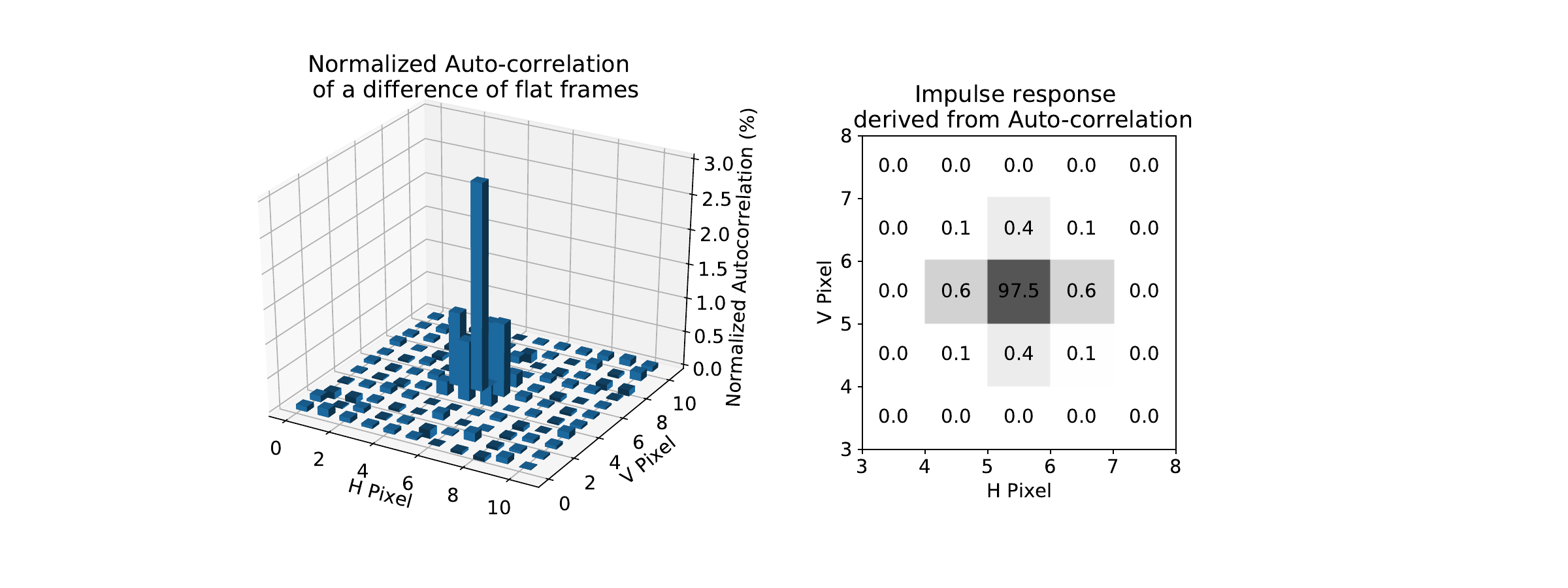}
    \end{center}
\caption{Left: Auto-correlation of the difference of two flat frames. The noise is reduced by averaging 48 $256\times256$ pixel patches. Right: Impulse response or IPC reconstructed from the Autocorrelation in low gain mode. In High-Gain mode, no significant inter pixel coupling is measured.}
\label{fig:IPC2DAutocorr}
\end{figure}

We extracted the impulse response from a difference of two flat frames using a 2D auto-correlation method.
To limit errors due to illumination non-uniformity and imperfections within the device itself, the area we used for processing was limited to well chosen $256\times256$ pixel subarrays.
This is only enough to provide a $\sim0.4\%$ error bar with our results according to [\citenum{Hirata:2019:10.1088/1538-3873/ab44f7}]. We performed a Monte Carlo simulation using fake shot-noise limited frames with no IPC to confirm this error bar. Forty-eight pairs of frames were then combined to reduce the errors. The normalized kernel of the inferred impulse response is shown in Figure~\ref{fig:IPC2DAutocorr}. 

\begin{figure}[h]
    \begin{center}
       \includegraphics[width=165mm]{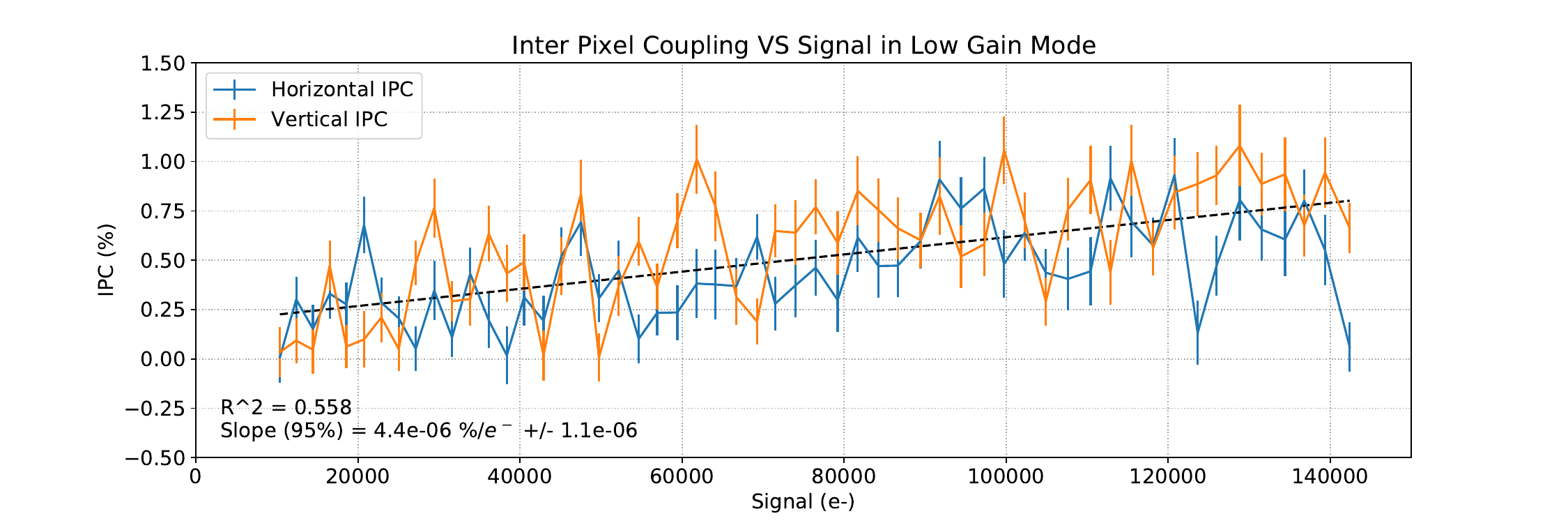}
    \end{center}
\caption{Horizontal and Vertical Correlation of first neighbors in low gain mode.The error reported on the slope justifies the linear fit versus a constant.}
\label{fig:IPCvsSignal}
\end{figure}

We reproduced this analysis for a range of illumination levels (Figure~\ref{fig:IPCvsSignal}). 
We define the horizontal (vertical) correlation as the ratio of the first horizontal (vertical) neighbors over the central peak.
Here, the noise associated with each sample is evaluated by computing the standard deviation of the correlation coefficient of pixels far away from the peak. Both horizontal and vertical directions display a similar rate of IPC increase with signal, a characteristic associated with decreasing slope in the variance versus mean as signal increases. Both effects are typically associated with BFE but confirmation awaits spot projection tests in which the dependence of lateral charge diffusion on signal contrast can be measured directly. It is relevant to note here that the pixel clock was slowed down to 1.25\,kHz to eliminate any inter-channel coupling or bandwidth limit coming from the video acquisition chain.

\begin{figure}[h]
    \begin{center}
       \includegraphics[width=170mm]{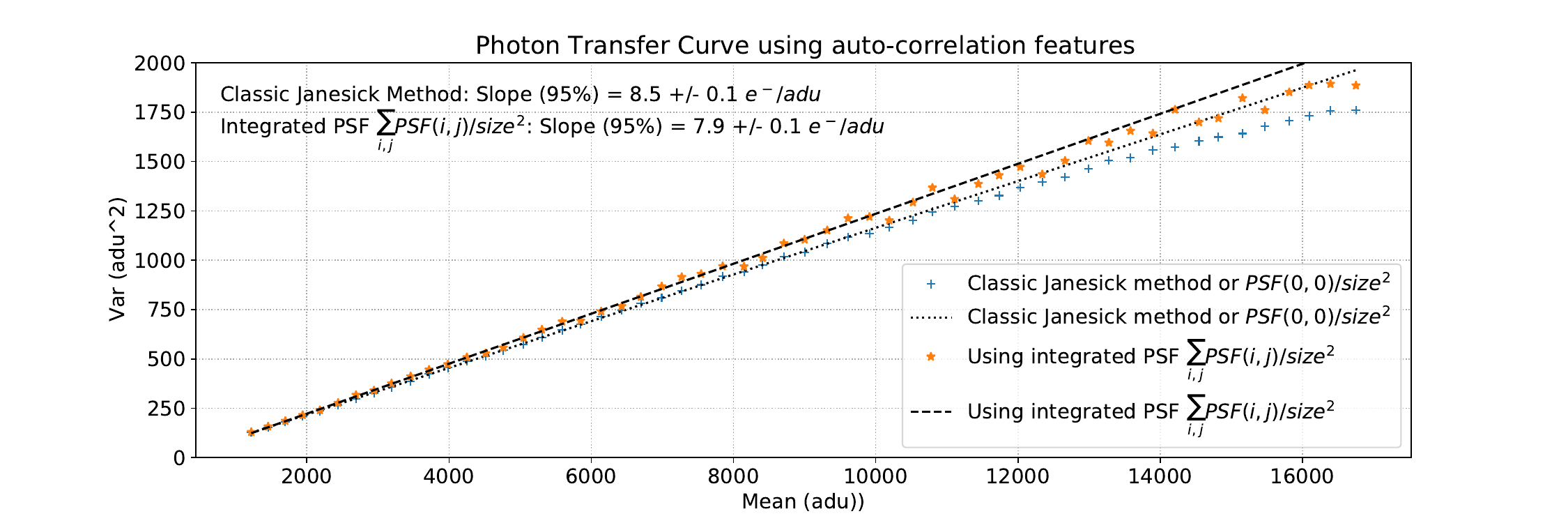}
    \end{center}
\caption{Photon Transfer curve obtained in Low-Gain mode by taking into account the pixel correlation. The total conversion gain drops by 6\% from $8.5\,\elec/\mathrm{ADU}$ to $7.9\,\elec/\mathrm{ADU}$  in low-gain mode.}
\label{fig:PTCusingIPC_LG}
\end{figure}

The classical variance-versus-mean method [\citenum{Janesick10.1117/3.725073}] assumes pixels are uncorrelated and therefore underestimates the shot noise of the incident photons due to the smoothing effect of the spatial correlation. Figure~\ref{fig:PTCusingIPC_LG} shows such a Photon Transfer Curve. A similar variance-versus-mean curve that integrates the power distributed in the wings of the impulse response reveals the true conversion gain. Before correction for correlations, the conversion gain is overestimated by $\sim6\%$. The corrected Conversion Gain has been used throughout this paper wherever units are in $\elec$.

\begin{figure}[h]
    \begin{center}
       \includegraphics[width=170mm]{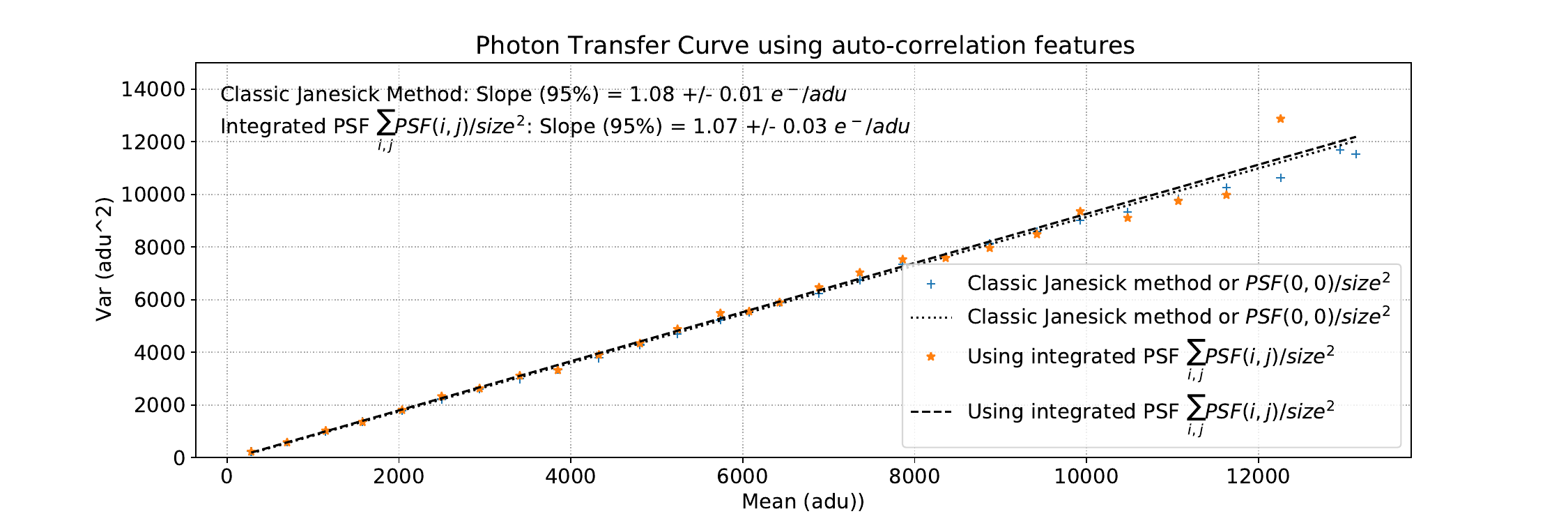}
    \end{center}
\caption{Photon Transfer curve obtained in High-Gain mode. The pixel correlation is insignificant.}
\label{fig:PTCusingIPC_HG}
\end{figure}

While inter-pixel coupling can be accurately measured in low-gain mode, no significant coupling was measurable in high-gain mode (Figure. \ref{fig:PTCusingIPC_HG}).

\subsection{Cosmic Ray Statistics}

 We have examined the distribution of cosmic rays incident on the detector from three 18 hour dark exposures of the CMOS detector in order to confirm that event rates and number of pixels affected are as expected. This information is useful for understanding how particle event sizes (track lengths) are likely to scale if thickness is changed, and to check that the detector package and AR coatings do not produce additional events. Bad pixels were masked to avoid spurious detections, but this masking did not include the noisiest pixels.

\begin{figure}[h]
    \begin{center}
       \includegraphics[width=150mm]{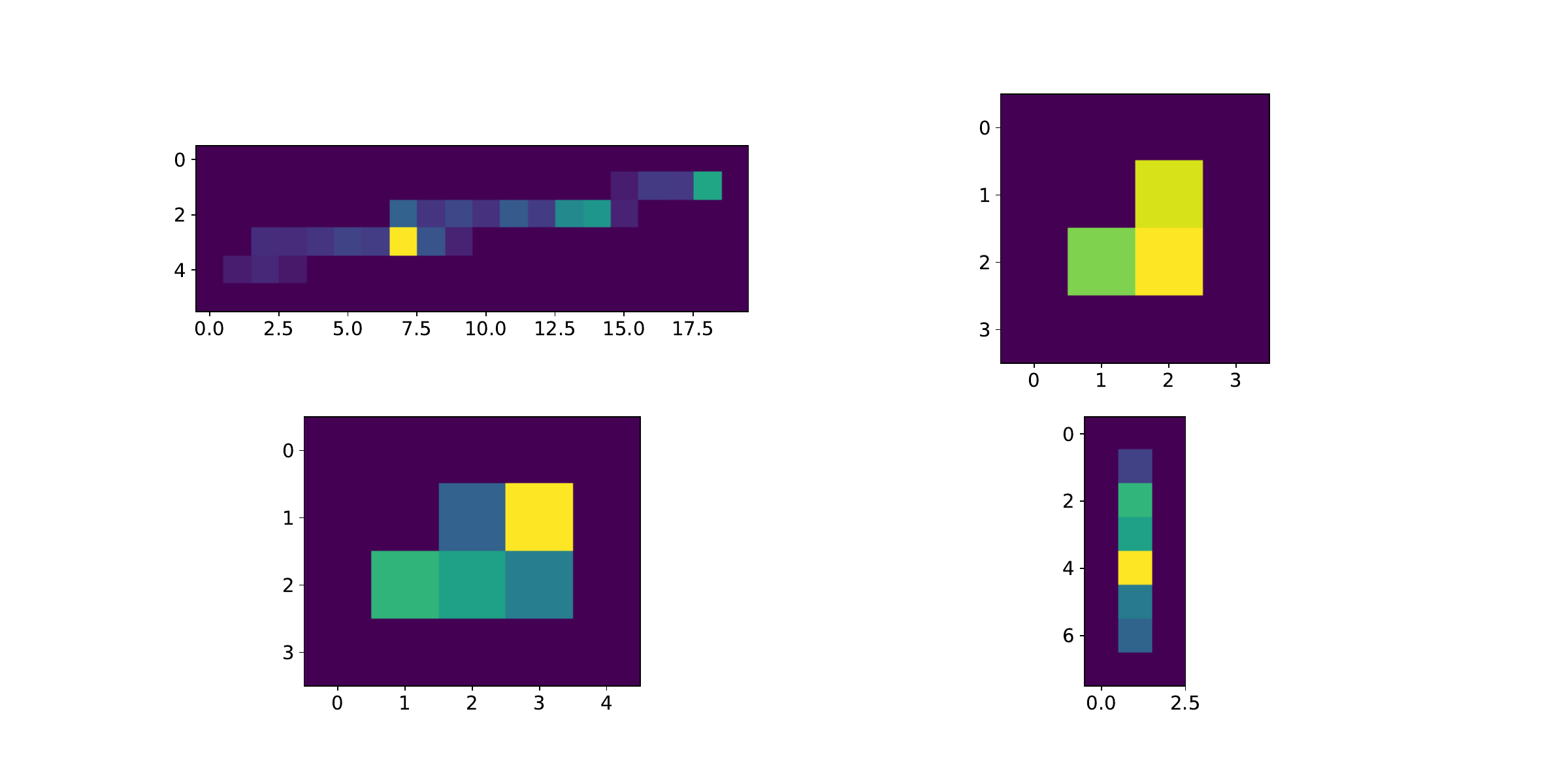}
    \end{center}
\caption{Example of various length cosmic ray events identified by finding contiguous regions above 50\,$\elec$ in dark frames.}
\label{fig:CREvent}
\end{figure}

Cosmic rays were identified as any contiguous region above a threshold set to 50 e-, well above the median read noise. This resulted in 31,877 events, or 1.45\,$\mathrm{CR/min/cm^2}$, which is 45\% higher than the expected 1\,$\mathrm{CR/min/cm^2}$  [\citenum{PhysRevD.86.010001}]. The excess might be caused by radiation from our concrete building. Figure~\ref{fig:CREvent} shows four examples of cosmic ray events identified with this method. We estimate the measured length distribution by taking the diagonal length of the bounding box of each cosmic ray. This crude method creates a large fractional error on the length of smaller cosmic rays.

\begin{figure}[h]
    \begin{center}
       \includegraphics[width=150mm]{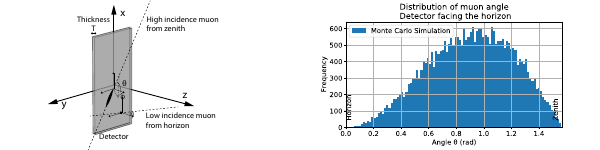}
    \end{center}
\caption{Left Figure: Reference diagram for incident angle $\theta$ and azimuthal angle $\phi$. \\ Right Figure: Expected angular distribution of cosmic rays hitting the detector. \\Muons and cosmic rays are preferentially coming from the zenith while our camera system is pointing towards the horizon. Note muons coming with a high incidence onto the detector leave a longer trace than those coming with a normal incidence.}
\label{fig:CR}
\end{figure}

For this experiment, the detector was facing the horizon. The incident angles of incoming cosmic rays, $\theta$ and $\phi$, were then calculated to follow the distribution of Equation \ref{Eq:MuonAngleDistribution} that accounts for the typical angular distribution of muons at the ground according to [\citenum{PhysRevD.86.010001}]. The resulting angle distribution is shown in Figure  \ref{fig:CR} (Right). A more in-depth analysis would need to account for radiation from the concrete room.

\begin{equation}
\begin{split}
    P(\theta,\phi) = cos^2(\theta-\pi/2))*cos(\theta)*cos(\phi)
    \label{Eq:MuonAngleDistribution}
\end{split}
\end{equation}

\noindent A diagram of the incident angle in reference to the cosmic ray and detector facing the horizon is shown in Figure \ref{fig:CR} (Left). The length $l$ of each Cosmic Ray track can then be related to the incident angle $\theta$ and $\phi$ and the thickness $T$ of the detector. 

\begin{figure}[h]
    \begin{center}
       \includegraphics[width=170mm]{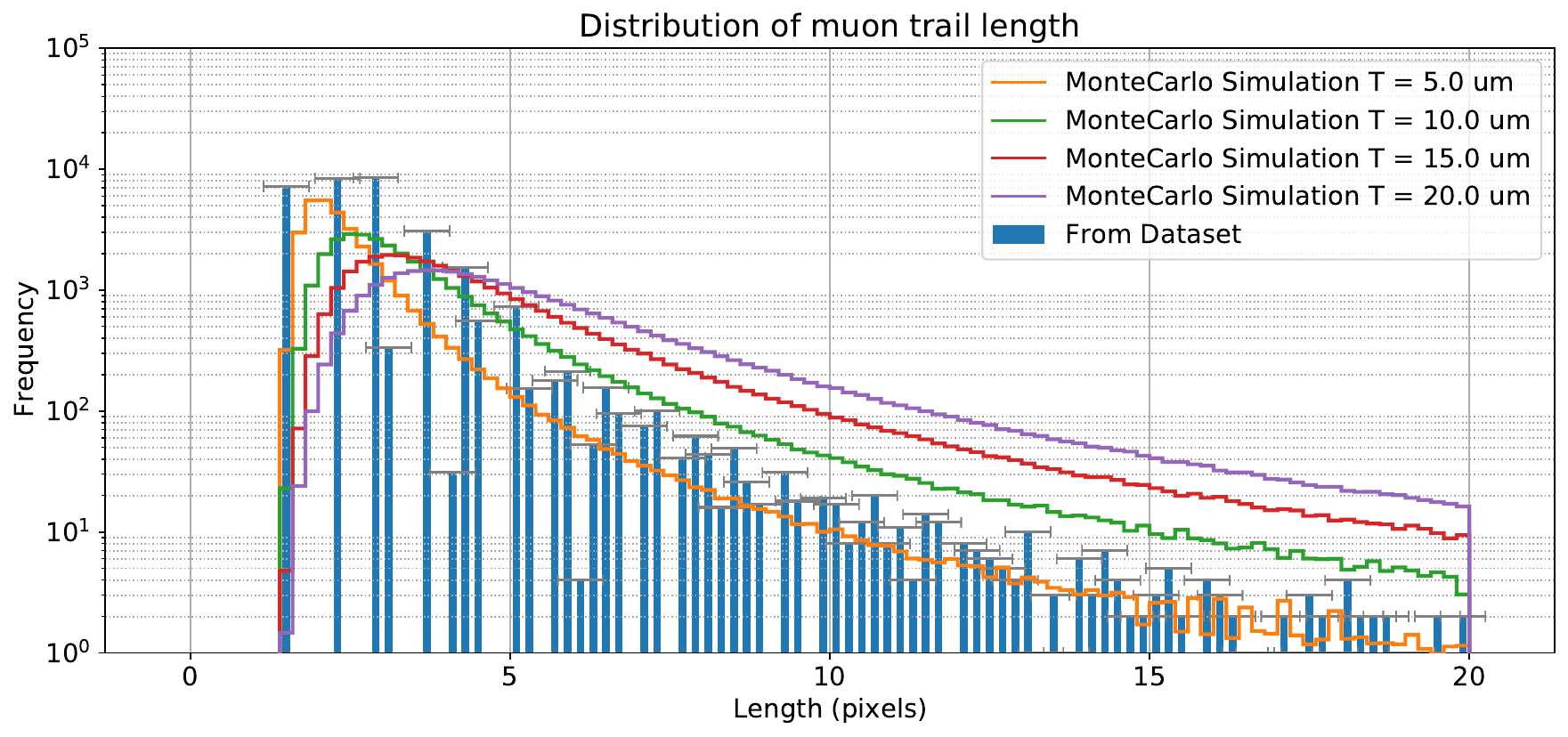}
    \end{center}
\caption{Comparison of cosmic ray track length distribution to Monte Carlo simulation. The error made when evaluating track length on our dataset is shown. To improve quality and smoothness of the simulated curves, trial numbers were boost up by 100x in each Monte Carlo simulations}.
\label{fig:LenDistrib_MonteCarlo}
\end{figure}

\begin{figure}[h]
    \begin{center}
       \includegraphics[width=170mm]{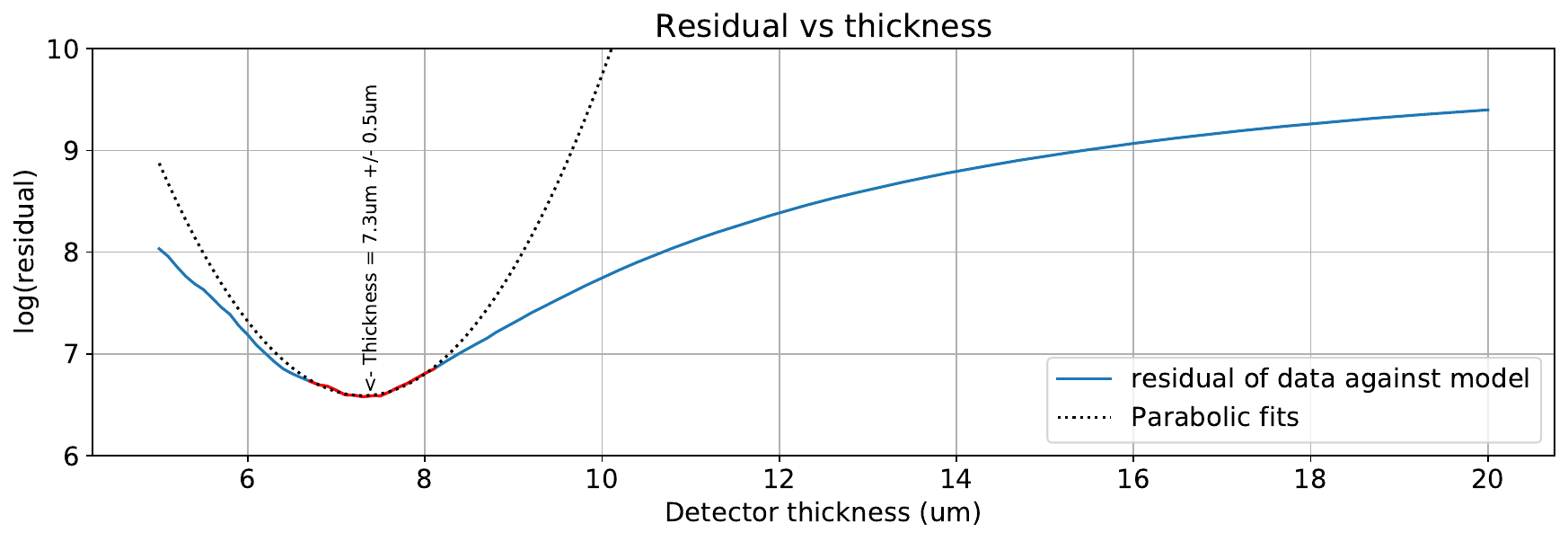}
    \end{center}
\caption{The residual is defined as  $ \epsilon = \sum{\frac{(data-MonteCarlo)^2}{MonteCarlo}}$ The minimum of this residual indicates the most likely detector's thickness. The error on this measurement is evaluated from the curvature of a fitted parabola around the minimum.}
\label{fig:ThicknessFit}
\end{figure}

Using this geometric model, we are able to use the cosmic ray length distribution to infer the  thickness of the detector, previously measured optically to be $7.0\pm0.3\mathrm{\mu}\mathrm{m}$. This was done by running a Monte Carlo simulation for the incident angles of cosmic rays, $\theta$ and $\phi$, drawn from the distribution in Figure \ref{fig:CR} (Right). We calculated the distribution in length of the simulated cosmic ray using the angle and thickness of the detector for a range of thicknesses from 5$\mathrm{\mu}\mathrm{m}$ to 20$\mathrm{\mu}\mathrm{m}$. The length distribution agrees with the expectation that shorter events are more likely, as longer cosmic ray tracks correlate with steeper incident angle. The mean track length is 3.6 pixels. The results of the Monte Carlo simulations are plotted alongside the measured length distribution shown in Figure~\ref{fig:LenDistrib_MonteCarlo} and compared using a least-square method to generate a graph showed in Figure~\ref{fig:ThicknessFit}, with the error calculated from a parabolic fit to the residual. The implied detector thickness is $7.3\pm0.5\mathrm{\mu}\mathrm{m}$, which overlaps with the direct optical measurement.

\noindent We determine from this length distribution that the total rate of pixels affected by cosmic rays is  4.1\,$\mathrm{pixel/min/cm^2}$ so fractional area contaminated is $4.1 \times 10^{-6}$ per minute.

\begin{figure}[h]
    \begin{center}
       \includegraphics[width=170mm]{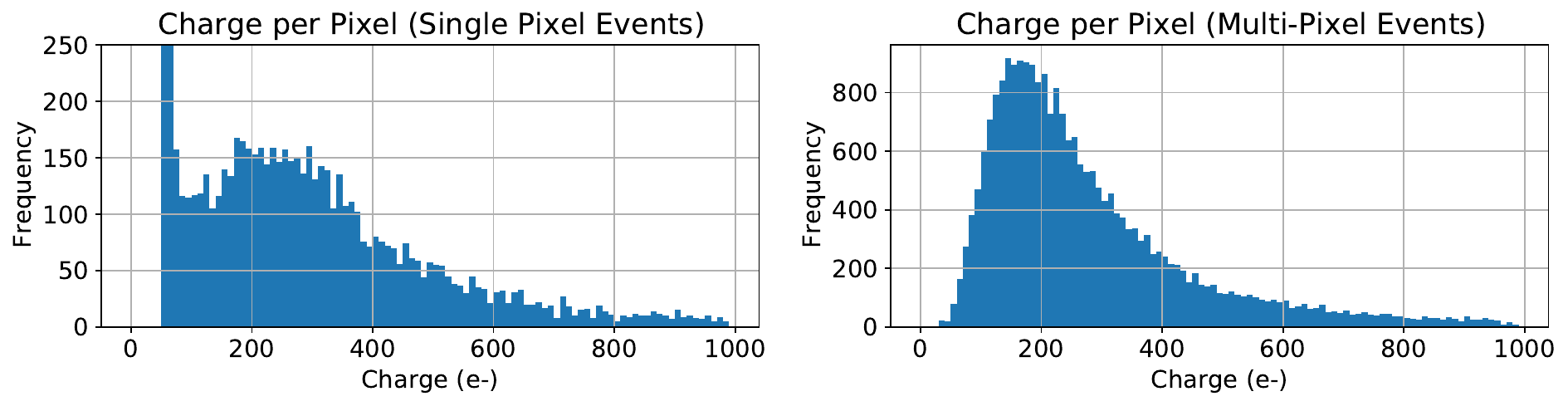}
    \end{center}
\caption{Cosmic Ray Charge Distribution for (left) all events and (right) multi-pixel events. The excess counts at lowest charge levels (at left) may be due to RTN and must be investigated further.}
\label{fig:CRDistributions}
\end{figure}

Figure~\ref{fig:CRDistributions} shows the charge distribution for all events and just multi-pixel events for comparison. We observe excess counts near the minimum charge deposition, suggesting that this is due to some pixels having sufficient Random Telegraph Noise (RTN) to exceed the CR detection threshold.  We assume the multi-pixel event distribution is the true distribution, and will return to masking pixels with excess RTN in a future work. The modal value of the multi-pixel charge distribution is 175\,$\elec$/pixel with estimated width 309\,$\elec$/pixel. 

\begin{figure}[H]
    \begin{center}
       \includegraphics[width=120mm]{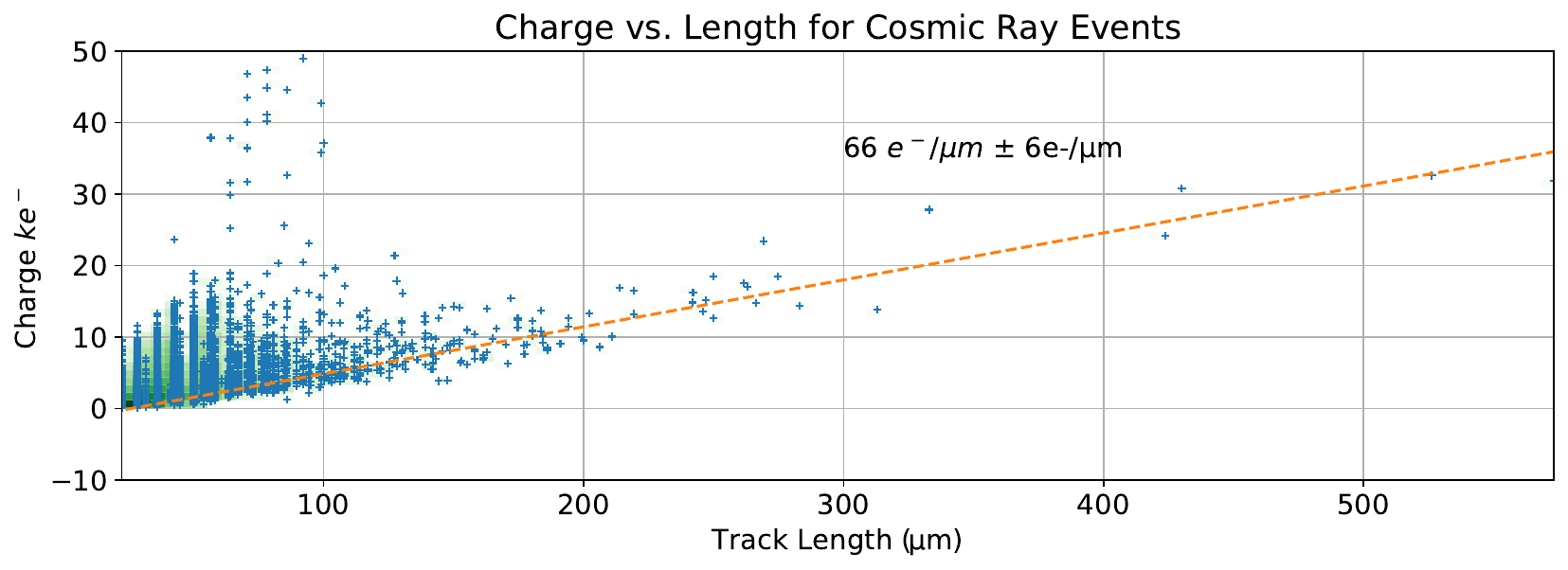}
    \end{center}
\caption{Charge vs. track length for cosmic ray events. The estimated slope is printed in $\mathrm{k\elec/\mathrm{\mu} m}$, with uncertainty estimated from the covariance matrix.}
\label{fig:ChargeTrackLength}
\end{figure}

98\% of cosmic rays are expected to be muons which create linear tracks with known deposition rate[\citenum{groom2002cosmic}]. To measure this deposition rate, we plotted the total charge deposited versus track length. This was measured in pixels, and thus is projected onto two dimensions rather than being the true three dimensional track length for each event.  The contribution of device thickness is significant only for track lengths less than $\sim100\,\mathrm{\mu} \mathrm{m}$.  Figure \ref{fig:ChargeTrackLength} shows an average $66\,\elec/\mathrm{\mu} \mathrm{m}$ charge deposition rate in the longest tracks region ($>100\mathrm{\mu} \mathrm{m}$), which is higher than the $27\,\elec/\mathrm{\mu} \mathrm{m}$ reported in [\citenum{groom2002cosmic}] for comparably thin silicon, but very similar to deposition rate measured with thick CCDs.

\section{Summary}
The SRI 4K$\times$2K described here is an ``engineering grade" device produced from a wafer that was part of a production run for the Europa Clipper mission. In spite of a post-processing error which produced several severe glow sites, dark current and read noise in the unaffected areas are excellent, rivalling the best CCDs. This is critical for UV spectroscopy and far UV imaging where sky background is extremely low.

In addition to being much more radiation tolerant than any CCD, this sensor offers higher exposure duty cycle, electronic shuttering, extended Dynamic Range (dual gain, dual exposure), and the ability to read subarrays (for guiding and/or photometry of very bright stars) while continuing long exposures on the remaining pixels. UVEX puts all of these features to good use. For example, several subarrays within each of its nine NUV sensors are located over bright stars which are read at 10 Hz to provide the precise guiding needed to keep a target centered on its 2 arcsec spectrograph slit, while also allowing photometry of these bright sources for cross-calibration against shallower legacy surveys.

The device exhibits excellent linearity.  Low correlation between adjacent pixels and the weak dependence of this correlation on signal  bodes well for future measurements of brighter-fatter effect.  The low thickness in relation to pixel size minimizes cosmic ray track length, while the narrowness of the tracks suggests low lateral charge diffusion, though this has not yet been carefully measured.

While this sensor meets all key requirements for UVEX, further work is desirable to quantify how far the photometric and astrometric accuracy can be pushed. Future tests will include  projection and scanning of thousands of sub-pixel spots to characterize charge diffusion, to investigate calibration systematics arising from errors in pixel boundary location (e.g. brighter-fatter effect), and to verify that photometry does not vary appreciably with sub pixel spot motion. More sensitive image persistence tests are also planned.

Meanwhile work is proceeding (at JPL) to validate coating performance, optimizing both in-band QE and out-of-band rejection. An essential part of this work is the validation of Quantum Yield (the production of more than one electron per photon) and the study of how QY statistics affect SNR as a function of fluence, a property shared by all silicon detectors.

\section{Acknowledgments}
We thank the SRI team for packaging multiple detectors and performing screening tests, and for their advice during our detector characterization effort. We acknowledge the support of JPL and Caltech's President and Director's Research and Development Fund (PDRDF Program) which supported the majority of this investigation. We also gratefully acknowledge the collaborative agreement between APL-SRI-JPL that made wafers available for this work.


\bibliography{report}   
\bibliographystyle{spiejour}   


\listoffigures
\listoftables

\end{spacing}
\end{document}